\documentclass[conference]{IEEEtran}
\usepackage{mathpazo}
\usepackage{times}
\usepackage{amsmath}
\usepackage{amsfonts}
\usepackage{latexsym}
\usepackage{amssymb}

\usepackage{upref}
\usepackage{theorem}
\usepackage{graphicx}
\usepackage{psfrag}
\usepackage{cite}

\usepackage{color}

\usepackage{algorithm}
\usepackage{algorithmicx}
\usepackage{algpseudocode}
\usepackage{verbatim}

\theoremstyle{plain}
\theorembodyfont{\normalfont\slshape}

\newtheorem{thm}{Theorem$\!$}
\newenvironment{theorem}
{\begin{thm}\hspace*{-1ex}{\bf.}}{\end{thm}}

\newtheorem{clm}[thm]{Claim$\!$}

\newtheorem{lem}[thm]{Lemma$\!$}
\newenvironment{lemma}{\begin{lem}\hspace*{-1ex}{\bf.}}{\end{lem}}

\newtheorem{prop}[thm]{Proposition$\!$}

\newtheorem{cor}[thm]{Corollary$\!$}
\newenvironment{corollary}{\begin{cor}\hspace*{-1ex}{\bf.}}{\end{cor}}

\newtheorem{defn}[thm]{Definition$\!$}
\newenvironment{definition}{\begin{defn}\hspace*{-1ex}{\bf.}}{\end{defn}}

\newtheorem{xmpl}[thm]{Example$\!$}
\newenvironment{example}{\begin{xmpl}\hspace*{-1ex}{\bf.}}{\hfill $\Box$ \end{xmpl}}

\newtheorem{cnstr}{Construction$\!$}

\newcounter{enumrom}
\renewcommand{\theenumrom}{(\roman{enumrom})}

\makeatletter
\renewcommand{\@endtheorem}{\endtrivlist}
\makeatother

\makeatletter
\renewcommand{\thefigure}{{\@arabic\c@figure}}
\renewcommand{\fnum@figure}{{\bf Figure\,\thefigure}}
\makeatother

\newcommand{\cA}{\mathcal{A}}
\newcommand{\cB}{\mathcal{B}}

\newcommand{\cF}{\mathcal{F}}

\newcommand{\cL}{\mathcal{L}}
\newcommand{\cM}{\mathcal{M}}

\newcommand{\cR}{\mathcal{R}}

\newcommand{\cX}{\mathcal{X}}
\newcommand{\cY}{\mathcal{Y}}

\newcommand{\mathset}[1]{\left\{#1\right\}}
\newcommand{\abs}[1]{\left|#1\right|}
\newcommand{\ceilenv}[1]{\left\lceil #1 \right\rceil}
\newcommand{\floorenv}[1]{\left\lfloor #1 \right\rfloor}
\newcommand{\fcenv}[1]{\left\lfloor #1 \right\rceil}
\newcommand{\parenv}[1]{\left( #1 \right)}
\newcommand{\sparenv}[1]{\left[ #1 \right]}
\newcommand{\bracenv}[1]{\left\{ #1 \right\}}
\newcommand{\rfrac}[2]{{}^{#1}\!/_{#2}}

\newcommand{\be}[1]{\begin{equation}\label{#1}}
\newcommand{\ee}{\end{equation}}

\renewcommand{\leq}{\leqslant}

\renewcommand{\geq}{\geqslant}

\renewcommand{\Bbb}{\mathbb}

\newcommand{\Tref}[1]{Theo\-rem\,\ref{#1}}

\newcommand{\Lref}[1]{Lem\-ma\,\ref{#1}}
\newcommand{\Cref}[1]{Co\-ro\-lla\-ry\,\ref{#1}}

\renewcommand{\Bbb}{\mathbb}

\newcommand{\N}{{\Bbb N}}

\newcommand{\R}{{\Bbb R}}
\newcommand{\Z}{{\Bbb Z}}

\DeclareMathOperator{\inc}{In}
\DeclareMathOperator{\out}{Out}
\DeclareMathOperator{\doo}{DO}
\DeclareMathOperator{\coo}{CO}

\newcommand{\ccap}{\mathsf{cap}}
\newcommand{\tcap}{\overline{\mathsf{cap}}}
\newcommand{\seq}[1]{\boldsymbol{#1}}
\newcommand{\iseq}[1]{\overline{\boldsymbol{#1}}}
\newcommand{\cyc}{\mathrm{cyc}}
\newcommand{\limup}[1]{\lim_{#1\rightarrow\infty}}
\newcommand{\limdown}[1]{\lim_{#1\rightarrow 0}}
\newcommand{\limsupup}[1]{\limsup_{#1\rightarrow\infty}}
\newcommand{\liminfup}[1]{\liminf_{#1\rightarrow\infty}}
\newcommand{\walk}[1]{\xrightarrow{#1}}

\newcommand{\bup}{b_{\scriptscriptstyle{\mathrm{U}}}}
\newcommand{\blo}{b_{\scriptscriptstyle{\mathrm{L}}}}
\newcommand{\lchop}{\cR}
\newcommand{\first}{\cL}
\newcommand{\perr}{P_{\mathrm{err}}}

\outer\def\proclaim #1. #2\par{\medbreak
 \noindent{\bf#1.\enspace}{\sl#2\par}%
 \ifdim\lastskip<\medskipamount \removelastskip\penalty55\medskip\fi}

\begin{document}

% paper title
\title{\textbf{Semiconstrained Systems}}

\author{
\IEEEauthorblockN{\textbf{Ohad Elishco}}
\IEEEauthorblockA{Electrical and Computer Engineering  \\
Ben-Gurion University of the Negev\\
Beer Sheva 8410501, Israel \\
\texttt{ohadeli@bgu.ac.il}\vspace{-2.0em}}
\and
\IEEEauthorblockN{\textbf{Tom Meyerovitch}}
\IEEEauthorblockA{Department of Mathematics  \\
Ben-Gurion University of the Negev\\
Beer Sheva 8410501, Israel \\
\texttt{mtom@math.bgu.ac.il}\vspace{-2.0em}}
\and
\IEEEauthorblockN{\textbf{Moshe Schwartz}}
\IEEEauthorblockA{Electrical and Computer Engineering  \\
Ben-Gurion University of the Negev\\
Beer Sheva 8410501, Israel \\
\texttt{schwartz@ee.bgu.ac.il}\vspace{-2.0em}}
}

\maketitle

\begin{abstract}
When transmitting information over a noisy channel, two approaches,
dating back to Shannon's work, are common: assuming the channel errors
are independent of the transmitted content and devising an
error-correcting code, or assuming the errors are data dependent and
devising a constrained-coding scheme that eliminates all offending
data patterns. In this paper we analyze a middle road, which we call a
semiconstrained system.  In such a system, which is an extension of
the channel with cost constraints model, we do not eliminate the
error-causing sequences entirely, but rather restrict the frequency in
which they appear.
%% By using this approach, we may later be able to decrease the rate
%% penalty, since a weaker error-correcting code is needed.

We address several key issues in this study. The first is proving
closed-form bounds on the capacity which allow us to bound the
asymptotics of the capacity. In particular, we bound the rate at which
the capacity of the semiconstrained $(0,k)$-RLL tends to $1$ as $k$
grows. The second key issue is devising efficient encoding and
decoding procedures that asymptotically achieve capacity with
vanishing error. Finally, we consider delicate issues involving the
continuity of the capacity and a relaxation of the definition of
semiconstrained systems.
\end{abstract}

%%%%%%%%%%%%%%%%%%%%%%%%%%%%%%%%%%%%%%%%%%%%%%%%%%%%%%%%%%%%%%%%%%%%%%%%
\section{Introduction}
%%%%%%%%%%%%%%%%%%%%%%%%
%%%%%%%%%%%%%%%%%%%%%%%%
%%%%%%%%%%%%%%%%%%%%%%%%

One of the most fundamental problems in coding and information theory
is that of transmitting a message over a noisy channel and attempting
to recover it at the receiving end. This is either when the
transmission is over a distance (a communications system), or over
time (a storage system). Two common solutions to this problem were
already described in Shannon's work \cite{Sha48a}. The first solution
uses an error-correcting code to combat the errors introduced by the
channel.  The theory of error-correcting codes has been studied
extensively, and a myriad of code constructions are known for a wide
variety of channels (for example, see
\cite{MacSlo78,LinCos04,Rot06,RicUrb08}, and the many references
therein). The second solution asserts that the channel introduces
errors in the data stream only in response to certain patterns, such
as offending substrings. It follows that removing the offending
substrings from the stream entirely will render the channel noiseless.
Schemes of this sort have been called constrained systems or
constrained codes, and they have also been extensively studied and
used (for example, see \cite{LinMar85,Imm04}, and the references
therein).

Both approaches are not free of cost. Error-correcting codes incur a
rate penalty, depending on the specific code used, and bounded by the
channel error model that is assumed. Constrained codes also impose a
rate penalty that is bounded by the capacity of the constrained
system.

The two solutions, one based on error-correcting codes and one based
on constrained codes, may be viewed as two extremes: while the first
assumes the errors are data independent, the second assumes the errors
are entirely data dependent. Since in the real world the situation may
not be either of the extremes, existing solutions may over-pay in rate.

The goal of this paper is to define and study semiconstrained systems
and their properties, as well as suggest encoding and decoding
procedures.

Arguably, the most famous constrained system is the $(d,k)$-RLL
system, which contains only binary strings with at least $d$ $0$'s
between adjacent $1$'s, and no $k+1$ consecutive $0$'s (see
\cite{Imm04} for uses of this system). In particular, $(0,k)$-RLL is
defined by the removal of a single offending substring, namely, it
contains only binary strings with no occurrence of $k+1$ consecutive
zeros, denoted $0^{k+1}$. Informally, a semiconstrained $(0,k)$-RLL
system has an additional parameter, $p\in [0,1]$, a real number. A
binary string is in the system if the frequency that the offending
pattern $0^{k+1}$ occurs does not exceed $p$. When $p=1$ this
degenerates into a totally unconstrained system that contains all
binary strings, whereas when $p=0$ this is nothing but the usual
constrained system, which we call a \emph{fully-constrained system}
for emphasis.

While the capacity of the semiconstrained $(0,k)$-RLL system is known
using the methods of \cite{MarRot1992}, the expression involves an
optimization problem that does not lend itself to finding other
properties of the system, such as the rate the capacity converges to
$1$ as $k$ grows. This rate of convergence is known when the system is
fully constrained \cite{SchVar11}. Additionally, the capacity is known
only in the one-dimensional case, whereas the general bounds may be
extended to the multi-dimensional case as well.

The first main contribution of this paper is establishing analytic
lower and upper bounds on the capacity of semiconstrained
$(0,k)$-RLL. These bounds are then used to derive the rate at which
the capacity of these systems converges to $1$ as $k$ grows, up to a
small constant multiplicative factor. The bounds extend previous
techniques from \cite{SchVar11} as well as employ large-deviations
theory. These bounds are also extended to the multi-dimensional case.

This paper is not motivated or limited solely by the case of a single
offending substring. We can define multiple offending substrings, each
equipped with its own limited empirical frequency. Indeed, with a
proper set of semiconstraints, variants such as DC-free RLL are
possible (see \cite{Kur11} and references therein).  Another
motivating example is the system of strings over $\Z_q$ where the
offending substring is $q-1,0,q-1$. In the case of multi-level flash
memory cells, inter-cell interference is at its maximum when three
adjacent cells are at the highest, lowest, and then highest charge
levels possible \cite{BerBir11}. By adjusting the amount of such
substrings we can mitigate the noise caused by inter-cell
interference. Further restrictions, such as the requirement for
constant-weight strings (see the recent \cite{KaySie14}) correspond to
a semiconstrained system.

Although coding schemes for some of these systems exist, they are
ad-hoc and tailored for each specific case, as in \cite{Kur11} and
\cite{KaySie14}. A more general coding scheme exists \cite{KhaNeu1996}
for a channel with cost constraints model. However, it is not optimal,
and it addresses only scalar cost functions, which is a different
model than the semiconstrained systems we study.

The second main contribution of this paper is a general explicit
encoding and decoding scheme. This coding scheme is based on the
theory of large deviations, and it asymptotically achieves capacity,
with a vanishing failure probability as the block length grows. To
that end, we also define and study a relaxation of semiconstrained
systems, allowing us to address the issue of the existence of the
limit in the definition of the capacity, as well as the continuity of
the capacity.

%% Although the capacity of a semiconstrained system with a single
%% offending word has been previously calculated, our main contribution
%% is emphasizing some delicate issues such as non-continuity and
%% overcoming those issues, finding an analytic expression for bounds on
%% the capacity, and calculating the rate in which the capacity goes to
%% $1$ as $k$ grows in the $(0,k)$ semicinstrained system.

%% We also prove an interesting connection with the theory of formal
%% languages. It is well known that fully-constrained system that are
%% defined by a finite set of forbidden substrings form regular languages
%% and are accepted by finite-state automata. In the case of
%% semiconstrained systems, even with a single offending substring, for
%% all choices of $p\in (0,1)$, $p\in\Q$ a rational number, we show the
%% semiconstrained system is never a regular language, but rather a
%% context-free language.

We would like to highlight some of the main differences between this
paper and previous works. In \cite{KhaNeu1996,KarNeuKha1988} the
capacity of channels with cost constraints is investigated. Such
channels define a scalar cost function that is applied to each
sliding-window $k$-tuple in the transmission. The admissible sequences
are those whose average cost per symbol is less than some given scalar
constraint.  In our paper, however, we investigate sequences with a
cost function which can control \emph{separately} the appearance of
any unwanted word (not necessarily of the same length).

The more general framework we study is similar to that of
\cite{MarTun1990,AlgMar1992,MarRot1992}. In \cite{MarTun1990} some
embedding theorems and results concerning the entropy of a
weight-per-symbol shift of finite type are presented, where the weights
are given by functions which take values in $\R^d$. In
\cite{AlgMar1992}, some large-deviation theorems are proved for
empirical types of Markov chains that are constrained to thin sets. A
thin set is a set whose convex hull has a strictly lower dimension
(which means it has an empty interior topologically). We also mention
\cite{MarRot1992}, in which an improved Gilbert-Varshamov bound for
fully-constrained systems is found. Thus, \cite{MarRot1992} studies
certain semiconstrained systems as means to an altogether different
end. Using these works, the exact capacity of semiconstrained systems,
as defined in this paper, may be calculated. However, key issues we
address are not covered by these papers, including the rate of
convergence of the capacity, the existence of the limit in the
capacity definition, and continuity of the capacity.

Finally, in \cite{KhaNeu1996}, coding for channels with cost
constraints is investigated. However, the main focus is given to
functions with a scalar cost on symbols, whereas the model we study in
this paper is different. The proposed coding scheme of
\cite{KhaNeu1996} is based on the state-splitting algorithm and is not
optimal in general.

The paper is organized as follows. In Section \ref{sec:defs} we give
the basic definitions and the notation used throughout the paper. We
also cite some previous work and derive some elementary
consequences. In Section \ref{sec:wscs} we introduce a relaxation
called weak semiconstrained systems, and study issues involving
existence of the limit in the capacity definition as well as
continuity of the capacity. In Section \ref{sec:bounds} we present an
upper and a lower bound on the capacity of the $(0,k,p)$-RLL
semiconstrained system, as well as bound the capacity's rate of
convergence as $k$ grows.  Section \ref{sec:encdec} is devoted to
devising an encoding and decoding scheme for weak semiconstrained
systems. We conclude in Section \ref{sec:conc} with a summary of the
results.

\section{Preliminaries}
\label{sec:defs}
%%%%%%%%%%%%%%%%%%%%%%%%%%%%%%%%%%%%
%%%%%%%%%%%%%%%%%%%%%%%%%%%%%%%%%%%%
%%%%%%%%%%%%%%%%%%%%%%%%%%%%%%%%%%%%

Let $\Sigma$ be a finite alphabet and let $\Sigma^*$ denote the set of
all the finite sequences over $\Sigma$. The elements of $\Sigma^*$ are
called \emph{words} (or \emph{strings}). The \emph{length} of a word
$\seq{\omega}\in\Sigma^*$ is denoted by $\abs{\seq{\omega}}$. Assuming
$\seq{\omega}=\omega_0\omega_1\dots\omega_{\ell-1}$, with
$\omega_i\in\Sigma$, a \emph{subword} (or \emph{substrings}) is a string
of the form $\omega_i\omega_{i+1}\dots\omega_{i+m-1}$, where $0\leq i\leq
i+m\leq \ell$.  For convenience, we define
\[
\seq{\omega}_{i,m}=\omega_{i}\omega_{i+1}\dots\omega_{i+m-1},
\]
i.e., $\seq{\omega}_{i,m}$ denotes the substring of $\seq{\omega}$
which is of length $m$ and is starting at the $i$th position. We say
$\seq{\omega}'$ is a \emph{proper subword} of
$\seq{\omega}\in\Sigma^*$ if $\seq{\omega}'$ is a subword of
$\seq{\omega}$, and $\abs{\seq{\omega}'}<\abs{\seq{\omega}}$.

Given two words, $\seq{\omega},\seq{\omega}'\in\Sigma^*$, their concatenation is
denoted by $\seq{\omega}\seq{\omega}'$. Repeated concatenation is denoted using
a superscript, i.e., for any natural $m\in\N$, $\seq{\omega}^m$ denotes
\[
\seq{\omega}^m=\seq{\omega}\seq{\omega}\dots\seq{\omega},\]
where $m$ copies of $\seq{\omega}$ are concatenated. As an example,
\[1(10)^301^20^3 = 1101010011000.\]

The following definition will be used later when defining
semiconstrained systems.

\begin{definition}
Let $\cF\subseteq\Sigma^*$ be a set of words.  We say that $\cF$ is
\emph{reduced} if $\seq{\omega}\in\cF$ implies no proper subword of
$\seq{\omega}$ is in $\cF$.
\end{definition}

For any two words $\seq{\tau},\seq{\omega}\in\Sigma^*$, let
$T(\seq{\tau},\seq{\omega})$ denote the frequency of $\seq{\tau}$ as a
subword of $\seq{\omega}$, i.e.,
\begin{equation}
\label{eq:defT}
T(\seq{\tau},\seq{\omega})=\frac{1}{\abs{\seq{\omega}}-\abs{\seq{\tau}}+1}\sum_{i=0}^{\abs{\seq{\omega}}-\abs{\seq{\tau}}}[\seq{\omega}_{i,|\seq{\tau}|}=\seq{\tau}].
\end{equation}
Here, $[A]$ denotes the Iverson bracket, having a value of $1$ if $A$
is true, and $0$ otherwise. If $\abs{\seq{\tau}}>\abs{\seq{\omega}}$ then we
define $T(\seq{\tau},\seq{\omega})=0$.

We are now ready to define semiconstrained systems.

\begin{definition}
Let $\cF\subseteq\Sigma^*$ be a finite reduced set of words, and let
$P\in [0,1]^{\cF}$ be a function from $\cF$ to the real interval
$[0,1]$. A \emph{semiconstrained system (SCS)}, $X{(\cF,P)}$, is the following
set of words,
\[X{(\cF,P)}=\mathset{ \seq{\omega}\in\Sigma^* : \forall\seq{\phi}\in\cF, T(\seq{\phi},\seq{\omega})\leq P(\seq{\phi}) }.\]
\end{definition}

When $\cF$ and $P$ are understood from the context, we may omit them
and just write $X$. We also define the set of words of length exactly
$n$ in $X{(\cF,P)}$ as
\[\cB_n(\cF,P) = X{(\cF,P)}\cap \Sigma^n.\]

\begin{example}
We recall that the $(d,k)$-RLL constrained system contains exactly the
binary words with at least $d$ $0$'s between adjacent $1$'s, and no
$k+1$ $0$'s in a row. Using our notation, after setting
$\Sigma=\mathset{0,1}$, the $(d,k)$-RLL constrained system is a
semiconstrained system $X{(\cF,P)}$, where
\[
\cF=\mathset{11,101,10^21,\dots,  10^{d-1}1,0^{k+1}},
\]
and $P(\seq{\phi})=0$ for all $\seq{\phi}\in\cF$.
\end{example}

An important object of interest is the capacity of an SCS.  We define
it as follows.
\begin{definition}
\label{capacity}
Let $X{(\cF,P)}$ be an SCS. The capacity of $X{(\cF,P)}$, which is denoted by
$\ccap(\cF,P)$, is defined as
\[
\ccap(\cF,P)=\limsup_{n\rightarrow \infty} \frac{1}{n}\log_2 \abs{\cB_n(\cF,P)}.
\]
\end{definition}

If we had a closed-form expression for $\abs{\cB_n(\cF,P)}$, we could
calculate the capacity of $(\cF,P)$. Following the same logic as
\cite{SchVar11}, we replace the combinatorial counting problem with a
probability-bounding problem. Assume $p_n$ denotes the probability
that a random string from $\Sigma^n$, which is chosen with uniform
distribution, is in $\cB_n(\cF,P)$. Then,
\[\abs{\cB_n(\cF,P)}=p_n\cdot \abs{\Sigma}^n,\]
and then
\begin{equation}
\label{eq:prob}
\ccap(\cF,P)=\log_2\abs{\Sigma}+\limsup_{n\rightarrow \infty} \frac{1}{n}\log_2 p_n.
\end{equation}
%% In the rest of the paper we shall either find the exact asymptotics of
%% $p_n$, or obtain bounds on it, thus, computing the exact capacity of a
%% semiconstrained system, or bounding it.

In certain cases, a different definition of semiconstrained systems is
helpful. The new definition has a cyclic nature. In a similar manner
to \eqref{eq:defT}, for any two words
$\seq{\tau},\seq{\omega}\in\Sigma^*$, let
$T^\cyc(\seq{\tau},\seq{\omega})$ denote the cyclic frequency of
$\seq{\tau}$ as a subword of $\seq{\omega}$, i.e.,
\[
T^\cyc(\seq{\tau},\seq{\omega})=\frac{1}{\abs{\seq{\omega}}}\sum_{i=0}^{\abs{\seq{\omega}}-1}[\seq{\omega}_{i,|\seq{\tau}|}=\seq{\tau}],
\]
where this time,
$\seq{\omega}_{i,m}=\omega_i\omega_{i+1}\dots\omega_{i+m-1}$, and the
indices are taken modulo $\abs{\seq{\omega}}$. We extend the
definitions of $X{(\cF,P)}$ and $\cB_n(\cF,P)$ to $X^\cyc{(\cF,P)}$
and $\cB_n^\cyc(\cF,P)$ in the natural way, by replacing $T$ with
$T^\cyc$.

Let $\bar{\Gamma}$ denote the closure of a set $\Gamma$, and let
$\Gamma^{\circ}$ denote its interior. Let $\cX$ be some Polish space
equipped with the Borel sigma algebra.

\begin{definition}
\label{ratefunction}
A \emph{rate function} $I$ is a mapping $I:\cX \rightarrow [0,\infty]$
such that for all $\alpha \in [0,\infty)$, the level set
$\phi_I(\alpha)=\mathset{x\in\cX : I(x)\leq \alpha}$ is a closed subset of
$\cX$. 
\end{definition}
\begin{definition}
\label{LDP}
Let $\mathset{\mu_n}$ be a sequence of probability measures. We say
that $\mathset{\mu_n}$ satisfies the large-deviation principle (LDP)
with a rate function $I$, if for every Borel set $\Gamma\subseteq\cX$,
\begin{align*}
-\inf_{x\in \Gamma^{\circ}} I(x)&\leq \liminf_{n\rightarrow \infty} \frac{1}{n}\log_2 \mu_n(\Gamma)\\ 
&\leq \limsup_{n\rightarrow\infty} \frac{1}{n}\log_2 \mu_n(\Gamma)\leq -\inf_{x\in \bar{\Gamma}} I(x).
\end{align*}
\end{definition}

Let $M_1(\Sigma)$ denote the space of all probability measures on some
finite alphabet $\Sigma$.
\begin{definition}
Let $\iseq{Y}=Y_0,Y_1,\dots$ be a sequence over some alphabet
$\Sigma$, and let $\seq{y}\in \Sigma^*$.  We denote by
$L_{n}^{\iseq{Y}}(\seq{y})$ the empirical occurrence frequency of the
word $\seq{y}$ in the first $n$ places of $\iseq{Y}$, i.e.
\[
L_{n}^{\iseq{Y}}(\seq{y})=T\parenv{\seq{y},\iseq{Y}_{0,n+\abs{\seq{y}}-1}}=\frac{1}{n}\sum_{i=0}^{n-1} \sparenv{{\iseq{Y}_{i,\abs{\seq{y}}}=\seq{y}}}.
\]
We denote by $L_{n,k}^{\iseq{Y}}\in M_1(\Sigma^k)$ the vector of
empirical distribution of $\Sigma^k$ in $\iseq{Y}$, i.e., for a
$k$-tuple $\seq{y}\in\Sigma^k$, the coordinate that corresponds to
$\seq{y}$ in $L_{n,k}^{\iseq{Y}}$ is $L_{n}^{\iseq{Y}}(\seq{y})$.
\end{definition}

Suppose $\iseq{Y}=Y_0,Y_1,\dots$ are $\Sigma$-valued i.i.d.~random
variables, with $q(\sigma)$ denoting the probability that
$Y_i=\sigma$, for all $i$. We assume that $q(\sigma)>0$ for all
$\sigma\in \Sigma$.  We denote by
$q(\sigma_0,\sigma_1,\dots,\sigma_{k-1})$ the probability of the
sequence $\sigma_0,\sigma_1,\dots,\sigma_{k-1}$. 
The following theorem connects the empirical distribution with the large-deviation principle.

\begin{theorem}{\cite[\S 3.1]{DemZei98}}
\label{ldtheo}
Let $\nu \in \cX=M_1(\Sigma^k)$, and let $\iseq{Y}=Y_0,Y_1,\dots$ be
$\Sigma$-valued i.i.d.~random variables, with $q(\sigma)>0$ denoting
the probability that $Y_i=\sigma$ for $\sigma\in\Sigma$. 
For every Borel set, $\Gamma\subseteq \cX$, define 
\[\mu_n\parenv{\Gamma}=\Pr\sparenv{L_{n,k}^{\iseq{Y}}\in \Gamma}.\]
Let us denote  by $\nu_1\in M_1(\Sigma^{k-1})$ the marginal of
$\nu$ obtained by projecting onto the first $k-1$ coordinates,
\[
\nu_1(\sigma_0,\dots,\sigma_{k-2})=\sum_{\sigma\in\Sigma}\nu(\sigma_0, \dots,
\sigma_{k-2},\sigma).
\]
Then the rate function, $I:\cX\rightarrow [0,\infty]$, governing the LDP of
the empirical distribution $L_{n,k}^{\iseq{Y}}$ with respect to $\Gamma$ is,
\[
I(\nu)=
\begin{cases}
\sum_{\seq{\sigma} \in\Sigma^k}\nu(\seq{\sigma}) \log_2 \frac{\nu(\seq{\sigma})}{\nu_1(\seq{\sigma}_{0,k-1})q(\sigma_{k-1})} & \text{$\nu$ is shift invariant,} \\
\infty & \text{otherwise},
\end{cases}
\]
where $\nu\in\cX= M_1(\Sigma^k)$ is shift invariant if 
\[
\sum_{\sigma\in \Sigma} \nu(\sigma,\sigma_1,\sigma_2,\dots, \sigma_{k-1})=\sum_{\sigma\in \Sigma} \nu(\sigma_1,\sigma_2, \dots, \sigma_{k-1},\sigma).
\]
\end{theorem}

In the context of an $X(\cF,P)$ SCS with
$\cF\subseteq \Sigma^k$, i.e., all the offending words are of equal length,
the set $\Gamma\subseteq M_1(\Sigma^k)$ takes on the following intuitive form,
\[ \Gamma=\mathset{ (p_{\seq{\phi}})_{\seq{\phi}\in \Sigma^k}\in M_1(\Sigma^k) : \forall \seq{\phi}\in \cF, p_{\seq{\phi}}\leq P(\seq{\phi})}.\]
In other words, $\Gamma$ contains all the vectors indexed by the
elements of $\Sigma^k$, such that each entry is a real number from
$[0,1]$, the entries sum to $1$, and each entry corresponding to an
offending word $\seq{\phi}\in\cF$ does not exceed $P(\seq{\phi})$.

Note that if $I$ is continuous and $\Gamma\subseteq\cX$ is such that
$\bar{\Gamma}=\Gamma^{\circ}$, then $\inf_{x\in
  \bar{\Gamma}}I(x)=\inf_{x\in \Gamma^{\circ}}I(x)$.  In that case the
limit of Definition \ref{LDP} exists, giving $\lim_{n\rightarrow
  \infty} \frac{1}{n}\log_2 \mu_n(\Gamma) = -\inf_{x\in \Gamma} I(x)$.

%% Note that Theorem \ref{ldtheo} implies that the limit in Definition
%% \ref{capacity} exists for some cases.  Since we are interested only in
%% shift-invariant measures we restrict our analysis to those measures.
%% Clearly, the limit exists for sets $\cF$ with equal-length offending
%% strings, i.e., $\cF\subseteq\Sigma^{k}$. \redcomment{ Why is the last
%%   sentence true, if at all?}

An important observation is the following. Assume
$\Sigma=\mathset{1,2,\dots,\abs{\Sigma}}$. Let us denote the vector of
probabilities for the symbols from $\Sigma$ by
$\seq{q}=(q(1),q(2),\dots,q(|\Sigma|))$. For a probability measure
$\mu\in M_1(\Sigma^{k-1})$ we define the probability measure
$\mu\otimes \seq{q}\in M_1(\Sigma^k)$ as follows.  The value of the
vector $\mu\otimes \seq{q}$ at the coordinate $i_1,i_2,\dots,i_k$ is
\[
(\mu\otimes \seq{q}) (i_1,\dots i_k)=\mu(i_1,\dots,i_{k-1})\cdot q(i_k).
\]
We now note that the rate function on the set of shift-invariant
measures, that governs the LDP of $L_{n,k}^{\iseq{Y}}$ as defined in
Theorem \ref{ldtheo}, can be written as
\[
I(\nu)=H(\nu|\nu_1\otimes \seq{q}),
\]
where $H(\cdot|\cdot)$ is the relative-entropy function. Since the
relative entropy is nonnegative and convex, and the set of
shift-invariant measures is closed and convex, we reach the following
corollary.
\begin{corollary}
\label{nonn}
The rate function governing the LDP of $L_{n,k}^{\iseq{Y}}$ defined in
Theorem \ref{ldtheo} is nonnegative and convex on the set of
shift-invariant measures $\nu\in M_1(\Sigma^k)$.
\end{corollary}

%% At this point we pause to formulate the strategy for computing the
%% capacity. By \eqref{eq:prob}, we need the limit of the probability
%% that a random sequence (chosen with uniform distribution) obeys the
%% constraints on the empirical distribution of the elements in $\cF$. We
%% therefore choose a sequence randomly by choosing each of its entries
%% i.i.d.~with uniform distribution over the alphabet.  By Definition
%% \ref{LDP}, the limit in question is bounded by the rate function,
%% whose form is given in Theorem \ref{ldtheo}.

%% This minimization problem is hard to solve in full generality.
%% Nevertheless, Theorem \ref{ldtheo} and Corollary \ref{nonn} give further
%% insight regarding some values of $P$ for which the capacity is equal
%% to $\log_2\abs{\Sigma}$.

Finally, the following corollary shows cases in which the constraints in
$P$ are redundant. 
\begin{corollary}
\label{cor:gencap}
Let $\cF\subseteq \Sigma^k$. 
If $P(\seq{\phi})\geq \abs{\Sigma}^{-k}$ for all
$\seq{\phi}\in\cF$, then $\ccap(\cF,P)=\log_2\abs{\Sigma}$.
\end{corollary}

\begin{IEEEproof}
Assume $\iseq{Y}=Y_0,Y_1,\dots$ is a sequence of i.i.d.~random
variables distributed uniformly (each symbol with probability
$\abs{\Sigma}^{-1}$).  Let $I$ be the rate function governing the LDP
of $L_{n,k}^{\iseq{Y}}$ as defined in Theorem \ref{ldtheo}. Consider
the shift-invariant measure $\nu\in M_1(\Sigma^k)$,
$\nu(\seq{\sigma})=\abs{\Sigma}^{-k}$, for all
$\seq{\sigma}\in\Sigma^k$.

Note that since $\cF$ is reduced, and since for all
$\seq{\phi}\in\cF$, $P(\seq{\phi})\geq \abs{\Sigma}^{-k}$, if the
probability for each $k$-tuple is $\abs{\Sigma}^{-k}$ then the
constraint holds.  Moreover, we obtain that $I(\nu)=0$ and by
Corollary \ref{nonn}, $\nu$ minimizes the rate function.  Let
$\Gamma\subseteq M_1(\Sigma^k)$ be the set associated with the
constraint, then
\begin{align*}
\ccap(\cF,P)&= \lim_{n\rightarrow\infty} \frac{1}{n}\log_2\parenv{\abs{\Sigma}^n\Pr[L_{n,k}^{\iseq{Y}}\in\Gamma]} \\
&= \log_2\abs{\Sigma}+ \lim_{n\rightarrow\infty} \frac{1}{n}\log_2\parenv{\Pr[L_{n,k}^{\iseq{Y}}\in\Gamma]} \\ 
&= \log_2\abs{\Sigma}-I(\nu) = \log_2\abs{\Sigma}.
\end{align*}
\end{IEEEproof}

\section{The Existence of the Limit in the Capacity Definition and
Weak Semiconstrained Systems}
\label{sec:wscs}
%%%%%%%%%%%%%%%%%%%%%%%%%%%%%%%%%%%%%%%%%%%%%%%%%
%%%%%%%%%%%%%%%%%%%%%%%%%%%%%%%%%%%%%%%%%%%%%%%%%
%%%%%%%%%%%%%%%%%%%%%%%%%%%%%%%%%%%%%%%%%%%%%%%%%
Let us consider the following two examples of binary semiconstrained
systems.

\begin{example}
Let $X(\cF,P)$ be an SCS with $\cF=\{0,1\}$ and
$P(0)=P(1)=\frac{1}{2}$.  Note that in this example the limit in the
definition of the capacity does not exist.  For an even number, $n$,
one can calculate $|\cB_n(\cF,P)|$ and obtain $\binom{n}{n/2}$ which
gives $\ccap(\cF,P)>0$.  For an odd $n$ we have that
$|\cB_n(\cF,P)|=0$.  It is easy to construct more examples in the same
spirit.
\end{example}

\begin{example}
Let $X(\cF,P)$ be an SCS with $\cF=\{0,1\}$ and $P(0)=r,\; P(1)=1-r$
where $r\in[0,1]$ is an irrational number.  We have that the possible
words are those with exactly an $r$-fraction of zeros and a
$(1-r)$-fraction of ones.  Since the capacity is defined on finite
words, for every $n$ we obtain $\cB_n(\cF,P)=\emptyset$, which implies
that $\ccap(\cF,P)=-\infty$.
\end{example}

These two examples are interesting because the first shows that the
limit in the definition of the capacity does not always exist, and the
second one shows that the capacity is not a continuous function of
$\seq{p}$.  That is, in the second example, from \Tref{ldtheo} we know
that for every $\epsilon>0$,
\[\limup{n} \frac{1}{n}\log_2 |\cB_n\parenv{\cF,(r+\epsilon,1-r+\epsilon)}|>0 \] 
exists. 
However, the second example shows that 
\begin{multline*} 
\limdown{\epsilon} \limup{n} \frac{1}{n}\log_2 |\cB_n\parenv{\cF,(r+\epsilon,1-r+\epsilon)}| \\ \neq \ccap\parenv{\cF,(r,1-r)}.
\end{multline*}

We therefore suggest a more relaxed definition of semiconstrained
systems.

\begin{definition}
Let $\cF\subseteq\Sigma^*$ be a finite reduced set of words, and let
$P\in [0,1]^{\cF}$ be a function from $\cF$ to the real interval
$[0,1]$.  A \emph{weak semiconstrained system (WSCS)},
$\overline{X}{(\cF,P)}$, is defined by
\[\overline{X}{(\cF,P)}=\mathset{ \seq{\omega}\in\Sigma^* : \forall\seq{\phi}\in\cF, T(\seq{\phi},\seq{\omega})\leq P(\seq{\phi})+\xi(\abs{\seq{\omega}}) },\]
where $\xi:\N\to\R^+$ is a function satisfying both $\xi(n)=o(1)$ and
$\xi(n)=\Omega(1/n)$. In addition, we define
\[\overline{\cB}_n(\cF,P)=\overline{X}{(\cF,P)}\cap \Sigma^n.\]
\end{definition}

We can think of $\xi(\abs{\seq{\omega}})$ as an additive tolerance to
the semiconstraints. The requirement that $\xi(n)=o(1)$ is in the
spirit of having the WSCS $\overline{X}$ ``close'' to the SCS $X$. In
the other direction, however, if we were to allow $\xi(n)=o(1/n)$,
then for large enough $n$, we would have gotten
$\cB_n=\overline{\cB}_n$, i.e., no relaxation at all. Thus, we require
$\xi(n)=\Omega(1/n)$.

The capacity of WSCS is defined in a similar fashion.
\begin{definition}
\label{capacity1}
Let $\overline{X}{(\cF,P)}$ be a WSCS. 
The capacity of $\overline{X}{(\cF,P)}$, which is denoted by $\tcap(\cF,P)$, is defined as
\[
\tcap(\cF,P)=\limsup_{n\rightarrow \infty} \frac{1}{n}\log_2 \abs{\overline{\cB}_n(\cF,P)}.
\]
\end{definition}

%We first note the for WSCS, the capacity and cyclic capacity are the same.
%
%\begin{lemma}
  %\label{lem:nocyciscyc}
  %Let $\overline{X}(\cF,P)$ be a WSCS and $\overline{X}^\cyc(\cF,P)$
  %its cyclic counterpart. Then,
  %\[\tcap(\cF,P)=\tcap^\cyc(\cF,P).\]
%\end{lemma}
%\begin{IEEEproof}
  %The proof is a consequence of the following simple observation: for
  %any any fixed $\seq{\tau}\in\Sigma^*$, and any arbitrarily long
  %$\seq{\omega}\in\Sigma^*$,
  %\[T(\seq{\tau},\seq{\omega})-T^\cyc(\seq{\tau},\seq{\omega})=O(1/\abs{\seq{\omega}}).\]
%\end{IEEEproof}

We show that under this definition of the capacity, the limit superior
is actually a limit.  Moreover, for cases such as the first example,
$\ccap(\cF,P)=\tcap(\cF,P)$. We do however note that weak
semiconstrained systems are not a generalization of fully-constrained
systems since the set of sequences in the latter does not contain any
word which belongs to $\cF$, while the former does.

In order to show that the limit in the definition of $\tcap$ exists we
need the following definition and theorems.  The first theorem shows
that it is possible to work over the set of shift-invariant measures
with the induced topology.
\begin{theorem}
\label{restriction}
Let $\cX=M_1(\Sigma^k)$ be the set of probability measures on
$k$-tuples and let $\cY=M_{\sigma}(\Sigma^k)\subseteq \cX$ be the set
of shift-invariant probability measures on $k$-tuples. If $\iseq{S}$
is a sequence of i.i.d.~symbols, denote
$\mu_n(\Gamma)=\Pr[L_{n,k}^{\iseq{S}}\in\Gamma]$.  Let $\mu_n|_{\cY}$
and $I|_{\cY}$ denote the restriction of $\mu_n$ and $I$ to $\cY$,
respectively.  If the sequence $\mathset{\mu_n}$ satisfies the LDP
with rate function $I$ then $\mu_n|_{\cY}$ satisfies the LDP with the
rate function $I|_{\cY}$.
\end{theorem} 

In order to prove Theorem \ref{restriction} we need some definitions
and a lemma.
\begin{definition}
A good rate function $I:\cX\to [0,\infty]$ is a rate function for which all the level sets are compact subsets of $\cX$.
\end{definition}

\begin{lemma}
The rate function governing the LDP of the empirical distribution
$L_{n,k}^{\iseq{Y}}$, as defined in \Tref{ldtheo}, is a good rate
function.
\end{lemma}

\begin{IEEEproof}
Recall that the rate function $I$ governing the LDP of the empirical
distribution $L_{n,k}^{\iseq{Y}}$, as defined in \Tref{ldtheo}, can be
written as the relative-entropy function.  Let us denote
$\ell=\abs{\Sigma}^k$.  The set $M_1(\Sigma^k)$ is isomorphic to a
closed and bounded subset of $[0,1]^{\ell}$ and hence, compact.  The
subset of shift-invariant measures in $M_1(\Sigma^k)$ is a closed
subset of $M_1(\Sigma^k)$ as a finite intersection of closed sets.
Every closed subset of a compact set is compact and therefore the set
of shift-invariant measures on $M_1(\Sigma^k)$ is compact.  Since $I$
is a rate function, the level sets are a closed subset of the
shift-invariant measures on $M_1(\Sigma^k)$ and hence compact.
\end{IEEEproof}

\begin{lemma}{\cite[Lemma 4.1.5]{DemZei98}}
\label{restriction2}
Let $\cX$ be a polish space and let $\cY\subseteq \cX$ be a
$G_{\delta}$-subset of $\cX$ (countable intersection of open sets in
$\cX$), equipped with the induced topology.  Let $\{\mu_n\}$ be finite
measures on $\cX$ such that $\mu_n(\cX\setminus\cY)=0$ for all $n\geq
1$ and let $I$ be a good rate function on $\cX$ such that
$I(x)=\infty$ for all $x\in\cX\setminus\cY$.  Let $\mu_n|_{\cY}$ and
$I|_{\cY}$ denote the restriction of $\mu_n$ and $I$ to $\cY$,
respectively.  Then, $I|_{\cY}$ is a good rate function on $\cY$ and
the following statements are equivalent.
\begin{enumerate}
\item The sequence $\{\mu_n\}$ satisfies the LDP with rate function $I$. 
\item The sequence $\{\mu_n|_{\cY}\}$ satisfies the LDP with rate function $I|_{\cY}$.
\end{enumerate}
\end{lemma}

We are now in a position to prove \Tref{restriction}.

\begin{IEEEproof}[Proof of \Tref{restriction}]
The shift-invariant measures on $M_1(\Sigma^k)$ form a
$G_{\delta}$-subset, and the probability of a sequence to have a
measure which is not shift invariant is $0$, i.e., $\mu_n(\cX\setminus
\cY)=0$ where $\cY$ is the set of shift-invariant measures.  Moreover,
by the definition of the rate function governing the LDP of the
empirical distribution, $L_{n,k}^{\iseq{Y}}$, we have $I(x)=\infty$
for $x\in\cX\setminus\cY$.  Thus, we can use \Lref{restriction2} which
allows us to restrict ourselves to the set of shift-invariant measures
with the induced topology, which is denoted by $M_{\sigma}(\Sigma^k)$.
\end{IEEEproof}

The set $\cF$ could contain words of various lengths, a fact that
sometimes complicates proofs.  In order to keep things as simple as
possible we would like to work with a set of forbidden words of the
same length, i.e., a set $\cF\subseteq \Sigma^k$ for some $k\in\N$. 
The next definition and theorem help us achieve this goal.

\begin{definition}
\label{caplim2}
Let $\cF\subseteq\Sigma^*$ be a finite reduced set of words.
Set $k=\max_{\seq{\phi}\in\cF}|\seq{\phi}|$ and define the operator
$f:M_1(\Sigma^k)\rightarrow [0,1]^{|\cF|}$ as follows.  Let $\cM$
be a $|\cF|\times |\Sigma|^k$ matrix, where for
$\seq{\phi}\in \cF$ and $\seq{\omega}\in\Sigma^k$, the
$(\seq{\phi},\seq{\omega})$ entry is given by
\[ 
\cM_{\seq{\phi},\seq{\omega}}=\sparenv{\seq{\omega}_{0,|\seq{\phi}|}=\seq{\phi}}.
\]
Then, for any $\nu\in M_1(\Sigma^k)$, we define $f(\nu)=\cM\nu$, where
$\nu$ is viewed as a vector indexed by $\Sigma^k$.
\end{definition}

\begin{example}
\label{ex:1}
Let $\cF=\mathset{1,100}$ thus $k=3$.  The matrix $\cM$ has two rows
and eight columns.  Each rows corresponds to a word from $\cF$: the
first row to $1$ and the second row to $100$. Each column corresponds
to a triple $000,001,010,\dots,111$.
\[
\cM= 
\begin{bmatrix} 
0 & 0 & 0 & 0 & 1 & 1 & 1 & 1 \\
0 & 0 & 0 & 0 & 1 & 0 & 0 & 0 
\end{bmatrix}.
\]
\end{example}

The key point we make is that by using $f^{-1}$ we are able to convert
constraints on a general set $\cF$ to constraints on $k$-tuples.  The
following theorem shows that the limit in the definition of the
capacity of the WSCS exists.
\begin{theorem}
\label{caplim}
Let $\overline{X}(\cF,P)$ be a WSCS, and 
\[\Delta=\bracenv{\nu\in [0,1]^{\cF} ~:~\forall\seq{\phi}\in\cF,\;\; \nu(\seq{\phi})\leq P(\seq{\phi})}.\]
Set $k=\max_{\seq{\phi}\in\cF} |\seq{\phi}|$, and let
$\cX=M_{\sigma}(\Sigma^{k})$ and $\cY=[0,1]^{\cF}$ (which is
isomorphic to $[0,1]^{|\cF|}$).  Define the linear function
$f:\cX\to\cY$ as in Definition \ref{caplim2}.
%For a $\seq{\omega}\in\cX$, the $\seq{\phi}\in\cF$ coordinate, \redcomment{Why is this notation for $f$ different from the one used above?}
%\[ 
%f(\seq{\omega})(\seq{\phi})=\frac{1}{k-|\seq{\phi}|+1}\sum_{i=0}^{k-|\seq{\phi}|}\sparenv{\seq{\omega}_{i,|\phi|}=\seq{\phi}}.
%\]
Then, if $f^{-1}(\Delta) \cap \cX \neq \emptyset$ the following equality holds (and the limit exists):
\[\tcap(\cF,P)=\limup{n} \frac{1}{n}\log_2 \abs{\overline{\cB}_n(\cF,P)}=1-\inf_{\mu\in f^{-1}(\Delta)} I(\mu), \]
provided the tolerance function for the WSCS satisfies $\xi(n)\geq
2\abs{\Sigma}^{k-1}n^{-1}$, and where $I$ is the rate function defined in
\Tref{ldtheo}.
\end{theorem}

Before proving \Tref{caplim} we need a technical lemma.
\begin{lemma}
\label{caplim3}
Let $\iseq{S}=S_0,S_1,\cdots$ be a sequence of $\Sigma$-valued
i.i.d.~random variables. Denote $\cY=[0,1]^{|\cF|}$, and let
$L^{\iseq{S}}_{n,\cF}$ denote the vector of empirical distribution of
words in $\cF$ in the first $n$ places of $\iseq{S}$, i.e., the
coordinate $L^{\iseq{S}}_{n,\cF}(\seq{\phi})$ that corresponds to
$\seq{\phi}\in\cF$ is $T(\seq{\phi},\iseq{S}_{0,n+|\seq{\phi}|-1})$.
Set $k=\max_{\seq{\phi}\in\cF} |\seq{\phi}|$.  Then, for all $n\geq
k$, and for every Borel set $\Delta\subseteq \cY$,
\[\Pr[L^{\iseq{S}}_{n,\cF}\in\Delta]=\Pr[L^{\iseq{S}}_{n,k}\in f^{-1}(\Delta)]. \] 
\end{lemma}
\begin{IEEEproof}
Let $L^{\iseq{S}}_{n,k}=v \in f^{-1}(\Delta)$, and
$L^{\iseq{S}}_{n,\cF}=u$.  We show that $u\in \Delta$.  Note that
$f(v)\in\Delta$ hence if $u=f(v)$ we are done. For all
$\seq{\phi}\in\cF$, we get,
\begin{align*} 
  f(v)(\seq{\phi})&=\sum_{\seq{\omega}\in\Sigma^k}\cM_{\seq{\phi},\seq{\omega}}v(\seq{\omega})\\
  &= \sum_{\seq{\omega}\in\Sigma^k} \sparenv{\seq{\omega}_{0,|\seq{\phi}|}=\seq{\phi}}T(\seq{\omega},\iseq{S}_{0,n+k-1}) \\
  &= \frac{1}{n} \sum_{\seq{\omega}\in\Sigma^k}\sum_{i=0}^{n-1} \sparenv{\iseq{S}_{i,k}=\seq{\omega}}\sparenv{\seq{\omega}_{0,|\seq{\phi}|}=\seq{\phi}} \\
	&= \frac{1}{n}\sum_{i=0}^{n-1}\sparenv{\iseq{S}_{i,|\seq{\phi}|}=\seq{\phi}} \\
	&= T(\seq{\phi},\iseq{S}_{0,n+\abs{\seq{\phi}}-1})=u(\seq{\phi}).
\end{align*}
The proof for the other direction is symmetric.
\end{IEEEproof}

\begin{example}
Let $\cF=\mathset{1,100}$ and $k=3$ as in Example \ref{ex:1}.  Consider
the sequence $\iseq{S}_{0,12}=101001101000$.  The empirical
distribution of triples in $\iseq{S}_{0,12}$ is shown in Table
\ref{tab1}.
\begin{table}[h!]
\caption{Empirical distribution of triples $L^{\iseq{S}}_{10,3}$ in the sequence $\seq{\omega}=101001101000\dots$.} \centering
\label{tab1}
\begin{tabular}{|c|c||c|c|}
\hline
\text{Triple} & \text{Distribution} & \text{Triple} & \text{Distribution} \\
\hline
\hline
$000$ & $\rfrac{1}{10}$ & $100$ & $\rfrac{2}{10}$ \\
$001$ & $\rfrac{1}{10}$ & $101$ & $\rfrac{2}{10}$ \\ 
$010$ & $\rfrac{2}{10}$ & $110$ & $\rfrac{1}{10}$ \\ 
$011$ & $\rfrac{1}{10}$ & $111$ & $0$ \\
\hline
\end{tabular}
\end{table}

Thus, 
\[ v=\parenv{\frac{1}{10},\frac{1}{10},\frac{2}{10},\frac{1}{10},\frac{2}{10},\frac{2}{10},\frac{1}{10},0}^T\]
and we have that 
\[ \cM v= \parenv{\frac{5}{10},\frac{2}{10}}^T.\] 
Indeed, $L^{\iseq{S}}_{10,\cF}(1)=T(1,\iseq{S}_{0,10})=\rfrac{5}{10}$ and
$L^{\iseq{S}}_{10,\cF}(100)=T(100,\iseq{S}_{0,12})=\rfrac{4}{12}$ and we obtain that
$L^{\iseq{S}}_{10,\cF}=\cM L^{\iseq{S}}_{10,3}$.
\end{example}

%\begin{lemma}
%\label{lem:equality}
%Let $\Delta\subseteq [0,1]^{|\cF|}$ and $\Gamma=f^{-1}(\Delta)$, where
%$f$ is defined as in Definition \ref{caplim2}.  Let $I$ be the rate
%function governing the LDP of the empirical distribution of $L_{n,k}$,
%as defined in \Tref{ldtheo}.  Set $\cX=M_1(\Sigma^k)$ and define
%$J(y)=\inf\bracenv{I(x):x\in\cX,\; y=f(x)}$.  Then,
%\[ \inf_{\seq{\nu}\in\Delta} J(\seq{\nu})=\inf_{\seq{\mu}\in \Gamma} I(\seq{\mu}). \]
%\end{lemma}
%
%\begin{IEEEproof}
%Let $r =\inf_{\seq{\nu}\in\Delta} J(\seq{\nu})$.  By definition, for
%every $\epsilon>0$ there exists $\seq{\nu}'\in\Delta$ such that
%$J(\seq{\nu}')<r+\frac{1}{2}\epsilon$.  Hence, by the definition of
%$J$, there exists $\seq{\mu}'\in f^{-1}(\seq{\nu}')$ such that
%$I(\seq{\mu}')<r+\epsilon$.  Since $\seq{\nu}'\in\Delta$ and
%$f(\seq{\mu}')=\seq{\nu}'$ we have that $\seq{\mu}'\in \Gamma$, which
%implies
%\[ \inf_{\seq{\nu}\in\Delta} J(\seq{\nu})\geq\inf_{\seq{\mu}\in \Gamma} I(\seq{\mu}). \]
%
%On the other hand, let $r =\inf_{\seq{\mu}\in \Gamma} I(\seq{\mu})$.
%For every $\epsilon>0$, there exists $\seq{\mu}'\in \Gamma$ for which
%$I(\seq{\mu}')<r+\epsilon$.  Since $\seq{\mu}'\in \Gamma$, we have
%$f(\seq{\mu}')\in\Delta$. Let us denote $\seq{\nu}'=f(\seq{\mu}')$.
%We obtain that
%\[ \inf_{\seq{\nu}\in\Delta} J(\seq{\nu})\leq J(\seq{\nu}')\leq I(\seq{\mu}')<r+\epsilon,\]
%which implies 
%\[ \inf_{\seq{\nu}\in\Delta} J(\seq{\nu})\leq\inf_{\seq{\mu}\in \Gamma} I(\seq{\mu}),\]
%and we have the desired result.
%\end{IEEEproof}

We are now ready to prove \Tref{caplim}.
\begin{IEEEproof}[Proof of \Tref{caplim}]
%% Let $\iseq{S}=S_0,S_1,\dots$ be a sequence of $\Sigma$-valued
%% i.i.d.~random variables, where $\Pr[S_i=\sigma]>0$ for all
%% $\sigma\in\Sigma$.  Denote by $L_{n,\cF}^{\iseq{S}}$ the vector of
%% empirical distribution of words from $\cF$ in
%% $(S_0,S_1,\dots,S_{n-1})$, i.e., the $\seq{\phi}\in\cF$ entry is
%% $T(\seq{\phi},(S_0,S_1,\dots,S_{n-1}))$ where we think of
%% $(S_0,S_1,\dots,S_{n-1})$ as cyclic.
We first note that $\Delta$ is closed and hence compact.  Let
$\{\Delta_n\}$ be the sequence
\[
\Delta_n=\mathset{ \nu\in [0,1]^{\cF} :
\forall\seq{\phi}\in\cF, \nu(\seq{\phi})\leq P(\seq{\phi})+\frac{2\abs{\Sigma}^{k-1}}{n} }.
\]
We let $\Gamma=f^{-1}(\Delta)$, and define the sequence
$\{\Gamma_n\}$ where for every $n$, $\Gamma_n=f^{-1}(\Delta_n)$, i.e.,
\begin{multline*}
\Gamma_n=\Bigg\{ \nu\in M_{\sigma}\parenv{\Sigma^k} : \\
\forall \seq{\phi}\in\cF,\;f(\nu(\seq{\phi}))\leq P(\seq{\phi})+\frac{2\abs{\Sigma}^{k-1}}{n} \Bigg\}. 
\end{multline*}
 
From \Tref{restriction} we can restrict ourselves to the
shift-invariant measures $M_{\sigma}(\Sigma^k)$.  Clearly, for every
$n$, $\Gamma_n$ is a closed and compact set, and
$\Gamma_{n}\to\Gamma$ when $n\to\infty$ (in the sense that for every open
neighborhood $U$ of $\Gamma$ there exists $N\in\N$ such that for all
$n>N$, $\Gamma_n\subseteq U$).  Moreover, $\bigcap_n \Gamma_n=\Gamma$.
Since $\Gamma$ is not empty and is closed, from LD theory we obtain
that for every $l\in\N$
\[ 
\limsupup{n} \frac{1}{n}\log\mu_n(\Gamma_n)\leq \limsupup{n} \frac{1}{n}\log\mu_n(\Gamma_l)\leq -\inf_{\nu\in\Gamma_l} I(\nu),
\] 
where $\mu_n(\cdot)=\Pr[L_{n,k}^{\iseq{S}}\in\cdot]$. 
Since the rate function is continuous we obtain, 
\[ 
\limsupup{n} \frac{1}{n}\log\mu_n(\Gamma_n)\leq \limup{l} \parenv{-\inf_{\nu\in\Gamma_l} I(\nu)} 
\stackrel{(a)}{=} -\inf_{\nu\in\Gamma} I(\nu)
\] 
where $(a)$ follows from the convergence of $\Gamma_n$ to $\Gamma$.

Now we argue that for every $n$, the set $\Gamma_n$ contains a
probability measure which belongs to the support of $\mu_n$, i.e.,
$\exists \; \omega_1\in\Gamma_n$ such that $\omega_1\in
L_{n,k}^{\iseq{S}}$.  Moreover, if $q_1\in\Gamma$ then we have
$\omega_1\in\Gamma_n$ such that $\omega_1\in L_{n,k}^{\iseq{S}}$ and
$\|q_1-\omega_1\|_{\infty}\leq \frac{2}{n}$ (see appendix for a
proof). Note that a tolerance of $2/n$ becomes, after applying $f$, a
tolerance of at most $2\abs{\Sigma}^{k-1}n^{-1}$ since the maximum sum of entries
in a row of $\cM$ is $\abs{\Sigma}^{k-1}$.  It is known
\cite{AlgMar1992} that
\[\frac{1}{n}\log\mu_n(\omega_1)=-I(\omega_1)+O(n^{-1}\log n).\]
For every $n\in\N$ 
\[\frac{1}{n}\log\mu_n(\Gamma_n)\geq -\inf_{\omega\in \Gamma_n\cap L^{\iseq{S}}_{n,k}}I(\omega)+O(n^{-1}\log n).\] 
Since the rate function is continuous and since $\|q_1-\omega_1\|_{\infty}\leq \frac{2}{n}$ we obtain
\[\frac{1}{n}\log\mu_n(\Gamma_n)\geq -\inf_{\omega\in \Gamma_n}I(\omega)+O(n^{-1}\log n)+o(1),\] 
which implies 
\[ \liminfup{n} \frac{1}{n}\log\mu_n(\Gamma_n)\geq -\inf_{\omega\in \Gamma}I(\omega).\] 
Therefore, 
\[ \limup{n}\frac{1}{n}\log \mu_n(\Gamma_n)=-\inf_{\omega\in \Gamma}I(\omega).\]
From \Lref{caplim3} and since 
\[
\mu_n(\Gamma_n)=\Pr[L_{n,k}^{\iseq{S}}\in f^{-1}(\Delta_n)]=\Pr[L_{n,\cF}^{\iseq{S}}\in \Delta_n]
\]
we have that 
\begin{align*}
\limup{n} \frac{1}{n}\log\mu_n(\Gamma_n)&=\limup{n} \frac{1}{n}\log \parenv{\Pr[L_{n,\cF}^{\iseq{S}}\in \Delta_n]} \\
&= -\inf_{\omega \in f^{-1}(\Delta)} I(\omega)
\end{align*}
as claimed.
\end{IEEEproof}

%% In the proof we take tolerance of $\frac{k2^k}{n}=O(1/n)$ instead of
%% $o(1)$.  While it has no effect on the upper bound $\limsupup{n}
%% \frac{1}{n}\log\mu_n(\Gamma_n)$, it affects the lower bound in the
%% since that we need to show that if $q_1\in\Gamma$ then we have
%% $\omega_1\in\Gamma_n$ such that $\omega_1\in L_{n,k}$ and
%% $\|q_1-\omega_1\|_{\infty}\leq \frac{2}{n}$.  This shows that indeed
%% we may limit ourselves to tolerance of $\frac{k2^k}{n}$.

The proof shows another important property, namely, the continuity of
the capacity as a function of $P$.  Note that the function $f$ is
continuous and the rate function is continuous when reduced to its
support.  Thus, if $P$ is not empty and $P+\seq{\epsilon}$ is not
empty,
$\limdown{\seq{\epsilon}}\tcap(\cF,P+\seq{\epsilon})=\tcap(\cF,P)$.

\section{Bounds on the Capacity of $(0,k,p)$-RLL SCS and Rate of Convergence}
\label{sec:bounds}
%%%%%%%%%%%%%%%%%%%%%%%%%%%%%%%%%%%%%%%%%%%%%%%%%%%%%%%%%%%%%%%%%%%%%%%%%
%%%%%%%%%%%%%%%%%%%%%%%%%%%%%%%%%%%%%%%%%%%%%%%%%%%%%%%%%%%%%%%%%%%%%%%%%
%%%%%%%%%%%%%%%%%%%%%%%%%%%%%%%%%%%%%%%%%%%%%%%%%%%%%%%%%%%%%%%%%%%%%%%%%

Although the capacity of semiconstrained systems may be found exactly
as a by-product of the methods described in \cite{MarRot1992}, it is
given as the result of an optimization problem. This does not allow us
to answer questions such as the rate of convergence of the capacity as
a function of its parameters.  In addition, the exact capacity
calculation applies only to the one-dimensional case.

In this section we study closed-form bounds on the capacity that allow
us to analyze the asymptotics of the capacity of semiconstrained
systems, and prove bounds on the capacity in the multi-dimensional
case. To that end, we focus on the family of semiconstrained
$(0,k)$-RLL, since it is defined by a single offending string of
$0^{k+1}$. While the results are specific to this family, we note that
some of them may be extended to general semiconstrained
systems.

For reasons that will become apparent later, we conveniently invert
the bits of the system, and define the $(0,k)$-RLL constrained system
as the set of all finite binary strings not containing the $1^{k+1}$
substring. We therefore consider the semiconstrained system $X(\cF,P)$
defined by
\[\cF=\mathset{1^{k+1}},\qquad\qquad P(1^{k+1})=p,\]
for some real constant $p\in [0,1]$. We call this semiconstrained
system the \emph{$(0,k,p)$-RLL SCS}, and throughout this section we
denote its capacity by $C_{k,p}$. Thus, $C_{k,0}$ denotes the capacity
of the fully-constrained $(0,k)$-RLL system.

In the case of fully-constrained $(0,k)$-RLL, the asymptotics of the
capacity, as $k$ tends to infinity, are well known. It was mentioned
in \cite{KatZeg99}, that
\[ 1 - C_{k,0} = \frac{\log_2 e}{4\cdot 2^k}(1+o(1)).\]
It was later extended in \cite{SchVar11} to the multi-dimensional case,
where it was shown that
\[ 1 - C_{k,0}^{(D)} = \frac{D\log_2 e}{4\cdot 2^k}(1+o(1)),\]
where $C_{k,0}^{(D)}$ denotes the capacity of the fully-constrained
$D$-dimensional $(0,k)$-RLL.

Our analysis will proceed by proving a lower and an upper on the
capacity of $(0,k,p)$-RLL SCS, and then analyzing it when
$k\to\infty$. We note that due to Corollary \ref{cor:gencap} we must
also take $p\leq \frac{1}{2^{k+1}}$ or else the capacity is exactly
$1$.

\subsection{An Upper Bound on the Capacity of $(0,k,p)$-RLL SCS}
%%%%%%%%%%%%%%%%%%%%%%%%
%%%%%%%%%%%%%%%%%%%%%%%%
%%%%%%%%%%%%%%%%%%%%%%%%

In order to obtain an upper bound on the capacity of $(0,k,p)$-RLL SCS
we employ a bound by Janson \cite{Jan90}. Consider an index set, $Q$,
and a set $\mathset{J_i}_{i\in Q}$ of independent random indicator
variables.  Let $\cA$ be a family of subsets of $Q$, namely,
$\cA\subseteq 2^Q$.  We define the random variables
\[S=\sum_{A\in \cA}I_A, \qquad\qquad I_A=\prod_{i\in A}J_i.\]
Moreover, we define
\[p_A=E[I_A], \quad
\lambda= E[S]=\sum_A p_A,\quad
\delta=\frac{1}{\lambda} \sum_{A\sim B}E[I_A I_B],
\]
where, for $A,B\in \cA$, we write $A\sim B$ if $A\cap B\neq \emptyset$
and $A\neq B$.

\begin{theorem}{\cite[Theorem 1]{Jan90}}
\label{janson}
If $\eta$ is an integer such that $0\leq \eta \leq \lambda$, then 
\[
\Pr\sparenv{S\leq \eta}\leq \parenv{ \sqrt{2\pi(\eta+1)}\frac{\lambda^{\eta}}{\eta!}e^{-\lambda}}^{\frac{1}{1+\delta}}.
\]
\end{theorem}

\begin{lemma} 
\label{lem:cyc}
For the $(0,k,p)$-RLL SCS, let
\[\cF=\mathset{1^{k+1}}, \qquad\qquad P(1^{k+1})=p\in[0,1].\]
Then
\[\ccap(X^\cyc(\cF,P))=\ccap(X(\cF,P)).\]
\end{lemma}

\begin{IEEEproof}
For any $\seq{\omega}\in \cB^\cyc_n$, $n\geq k+1$, by definition,
$\seq{\omega}\seq{\omega}_{0,k}\in \cB_{n+k}$. Thus,
$|\cB^\cyc_n|\leq\abs{\cB_{n+k}}$ for all $n\geq k+1$. In the other
direction, we note that for any $\seq{\omega}\in \cB_{n-1}$, one can easily
verify that $\seq{\omega}0\in \cB^\cyc_n$, for all $n\geq k+2$. It follows that,
for all $n\geq k+2$,
\[ \abs{\cB_{n-1}}\leq \abs{\cB^\cyc_n} \leq \abs{\cB_{n+k}}.\]
Taking the appropriate limits required by the definition of the
capacity, we prove the claim.
\end{IEEEproof}

Before stating the upper bound on the capacity of $(0,k,p)$-RLL SCS,
we explain briefly how Theorem \ref{janson} is going to be used.  We
conveniently set the index set of a string of length $n$ to be
$Q=\mathset{0,1,\dots, n-1}$. For $(0,k,p)$-RLL SCS we define the
family of subsets of $Q$,
\[\cA=\mathset{\mathset{i, i+1,\dots, i+k} : 0\leq i<n},\]
where the coordinates are taken modulo $n$. Setting $\eta=pn$, we have
that $\Pr[S\leq pn]$ is the probability that a sequence of length $n$
obeys the cyclic $(0,k,p)$-RLL SCS, for some integer $0\leq pn\leq \lambda$.
For this reason, Corollary \ref{cor:gencap} implies that
in case $\eta\geq\lambda$ the capacity is $\log_2 2=1$.

\begin{theorem}
\label{upper_bound}
For $0< p\leq \frac{1}{2^{k+1}}$, the capacity of the $(0,k,p)$-RLL
SCS is bounded by
\begin{align*}
C_{k,p} & \leq 1- \frac{1}{3-2^{-k+1}}\parenv{\frac{\log_2 e}{2^{k+1}}
+p(k+1)-p\log_2{\frac{e}{p}}}.
\end{align*}
\end{theorem}

\begin{IEEEproof}
Assume a sequence of $n$ bits are randomly chosen
i.i.d.~Bernoulli$(1/2)$.  It follows that
\[
\lambda = E[S]= \sum_A p_A = \frac{n}{2^{k+1}}.
\]

For each $A\in \cA$ there are exactly $2k$ sets, $B_i\in\cA$,
$i=0,1,\dots, 2k-1$, such that $A\sim B_i$.  If $\abs{A\cap B_i}=t$
then $E[I_AI_{B_i}]=\frac{1}{2^{2(k+1)-t}}$.  Hence,
\begin{align*}
\delta &= \frac{1}{\lambda}\sum_{A\in \cA}\sum_{B\sim A}E[I_AI_B] 
= \frac{1}{\lambda}\sum_{A\in \cA}\sum_{t=1}^{k}2\frac{1}{2^{2(k+1)-t}} \\
&= \frac{1}{\lambda}\sum_{A\in \cA}\frac{2^k-1}{2^{2k}}
= 2-\frac{1}{2^{k-1}}.
\end{align*}
Applying Theorem \ref{janson} yields
\begin{align*}
C_{k,p}&=\limsup_{n\rightarrow \infty}\frac{1}{n}\log_2{|B_n|} \\
&= \limsup_{n\rightarrow \infty}\frac{1}{n}\log_2{2^n\Pr[S\leq pn]} \\
&= 1+ \limsup_{n\rightarrow \infty}\frac{1}{n}\log_2{\Pr[S\leq \floorenv{pn}]} \\
&\stackrel{(a)}{\leq} 1+\lim_{n\rightarrow \infty} \frac{1}{n}\log_2 \parenv{ \sqrt{2\pi(\floorenv{pn}+1)}\frac{\lambda^{\floorenv{pn}}}{\floorenv{pn}!}e^{-\lambda}}^{\frac{1}{1+\delta}} \\
&= 1+ \lim_{n\rightarrow \infty} \frac{1}{(3-\frac{1}{2^{k-1}})n}\parenv{-\lambda \log_2{e} + \log_2{\frac{\lambda^{\floorenv{pn}}}{\floorenv{pn}!}}} \\
&= 1- \frac{\log_2 e}{(3-\frac{1}{2^{k-1}})2^{k+1}} \\
&\quad\  +\lim_{n\rightarrow \infty} \frac{1}{(3-\frac{1}{2^{k-1}})n}\log_2{\frac{n^{\floorenv{pn}}}{2^{\floorenv{pn}(k+1)} (\floorenv{pn}!)}} \\
&\stackrel{(b)}{=} 1- \frac{\log_2 e}{(3-\frac{1}{2^{k-1}})2^{k+1}} -\frac{p(k+1)}{(3-\frac{1}{2^{k-1}})} \\ 
&\quad\  + \frac{p}{(3-\frac{1}{2^{k-1}})}\log_2{\frac{e}{p}},
\end{align*}
where $(a)$ follows from Theorem \ref{janson} and from the existence
of the limit, and $(b)$ follows from Stirling's approximation.  Using
Lemma \ref{lem:cyc} we complete the proof.
\end{IEEEproof}

The same method can be applied for the $D$-dimensional $(0,k,p)$-RLL
SCS, extending the results of \cite{SchVar11}. We briefly define the
extension of SCS to the multi-dimensional case, and only sketch the
proof since it is similar to that of \cite{SchVar11}.

Define $[n]=\mathset{0,1,\dots,n-1}$, and let $\seq{e}_j$ be the $j$th
standard unit vector, containing all $0$'s, except the $j$th position
which is $1$. Assume $\Sigma$ is a finite alphabet, and
$\seq{\omega}\in\Sigma^{[n]^D}$ is an $n\times\dots \times n$
$D$-dimensional array over $\Sigma$. A substring of $\seq{\omega}$ is defined
as
\[\seq{\omega}_{\seq{i},m,\seq{e}_j}= \omega_{\seq{i}}\omega_{\seq{i}+\seq{e}_j}\dots \omega_{\seq{i}+(m-1)\seq{e}_j},\]
where $\seq{i}\in [n]^D$ is a $D$-dimensional index. We note that
$\seq{\omega}_{\seq{i},m,\seq{e}_j}$ is a one-dimensional string of
length $m$. We naturally extend $T^\cyc$ to the $D$-dimensional case
in the following manner: For $\seq{\omega}\in\Sigma^{[n]^D}$ and
$\tau\in\Sigma^*$, the frequency of $\tau$ as a cyclic substring of
$\seq{\omega}$ is defined as
\[T^\cyc(\seq{\tau},\seq{\omega})=\frac{1}{n^D}\sum_{j=0}^{D-1}\sum_{\seq{i}\in[n]^D} [\seq{\omega}_{\seq{i},\abs{\seq{\tau}},\seq{e}_j}=\seq{\tau}],\]
where indices are taken modulo $n$ appropriately.

The $D$-dimensional cyclic $(0,k,p)$-RLL SCS is defined as
\begin{align*}
  \cB_n^{\cyc,(D)}(1^{k+1},p) &= \mathset{ \seq{\omega}\in\mathset{0,1}^{[n]^D} :
    T^{\cyc}(1^{k+1},\seq{\omega})\leq p},\\
  X^{\cyc,(D)}(1^{k+1},p)&=\bigcup_{n} \cB_n^{\cyc,(D)}(1^{k+1},p).
\end{align*}
Its capacity is defined as
\[C_{k,p}^{(D)}=\limsup_{n\to\infty}\frac{1}{n^D}\log_2\abs{\cB_n^{\cyc,(D)}(1^{k+1},p)}.\]

We obtain the following upper bound on $C_{k,p}^{(D)}$:
\begin{theorem}
The capacity of the $D$-dimensional $(0,k,p)$-RLL SCS is bounded by
the following.
\[
C_{k,p}^{(D)}\leq 1-\frac{D\frac{\log_2(e)}{2^{k+1}}+p(k+1)-p\log_2\frac{De}{p}}{3-2^{-k+1}+2^{-k}(D-1)(k+1)^2}.
\]
\end{theorem}
\begin{IEEEproof}[Sketch of proof]
We use Janson's method, a direct calculation of the expected number of appearances of a sequence of $k+1$ ones, 
together with direct calculations of the value of $\delta$. 
We obtain that 
\begin{align*} 
\lambda&=\frac{Dn^D}{2^{k+1}},\\
\delta&= \frac{1}{\lambda}\sum_{A\in \cA}\sum_{B\sim A}E[I_AI_B]
= 2-\frac{1}{2^{k-1}}+\frac{(D-1)(k+1)^2}{2^{k}}.
\end{align*}
Placing $\lambda$ and $\delta$ in \Tref{janson} yields the wanted result.
\end{IEEEproof}

We return to the one-dimensional case. The upper bound of
\Tref{upper_bound} converges to $1$ as $k$ grows.  We now find the
rate of this convergence. To that end we prove a stronger upper bound
on the capacity, that does not have a nice form as Theorem
\ref{upper_bound} in the finite case, but does have a nice asymptotic
form. Note that $p$ must be a function of $k$ since $p\leq
\frac{1}{2^{k+1}}$.

\begin{theorem}
\label{th:upasbnd}
Let $c=\limup{k} \frac{p}{2^{-(k+1)}}$, where $c\in [0,1]$, and let
\[ \blo= \begin{cases}
\frac{3-\sqrt{1+8c}}{2}\log_2 e
-2c\log_2\parenv{\frac{1+4c+\sqrt{1+8c}}{8c}} & c>0, \\
\log_2 e & c=0.
\end{cases}
\]
Then, 
\[ 1-C_{k,p} \geq \frac{\blo}{2^{k+2}}(1+o(1)), \] 
where $o(1)$ denotes a function a of $k$ tending to $0$ as $k\to\infty$.
\end{theorem}
\begin{IEEEproof} 
Let $S'_A=I_{A}+\sum_{B\sim A}{I_{B}}$ and hence, given $I_A=1$,
$S'_A\in \mathset{1,2,\dots,2k+1}$.  We denote $\psi(t)=E[e^{-tS}]$.
For all $t\geq 0$, it was shown in \cite{Jan90} that
\begin{align}
-\frac{d}{dt}\ln \psi(t) &= \frac{1}{\psi(t)}\sum_{A\in\cA}E[I_Ae^{-tS}] \nonumber \\
\label{eq:z}
&\geq \sum_{A\in\cA}p_AE[e^{-tS'_A}|I_A=1].
\end{align}

While \cite{Jan90} bounded $z=E[e^{-tS'_A}|I_A=1]$, we proceed by
calculating it explicitly. Due to symmetry, $z$ does not depend on
$A$ or $n$ (for large enough $n$). Thus, \eqref{eq:z} becomes
\begin{equation}
\label{eq:deriv}
-\frac{d}{dt}\ln \psi(t) \geq z \sum_{A\in\cA}p_A = \lambda z.
\end{equation}
However, $z$ does depend on $k$ and $t$, which we will sometime
emphasize by writing $z(k,t)$.

We assume that the length of the sequence is at least $3k$, and
recall that we may consider sequences cyclically.  A tedious
calculation gives
\[
\Pr[S'_A=\ell|I_A=1]= 
\begin{cases}
\frac{\ell}{2^{\ell+1}}       & 1<\ell\leq k \\
\frac{2k+4-\ell}{2^{\ell+1}}  & k<\ell\leq 2k \\
\frac{4}{2^{\ell+1}}       & \ell=2k+1.
\end{cases}
\]
Thus,
\begin{align*}
z &= \sum_{j=1}^{k}{\frac{j}{2^{j+1}}e^{-tj}}+\sum_{j=k+1}^{2k}{\frac{2k+4-j}{2^{j+1}}e^{-tj}} + \frac{4}{2^{2k+2}}e^{-t(2k+1)}\\
&=\frac{2^{-2k}e^{-(1+2k)t}\parenv{e^{t}-2^ke^{(k+1)t}-1}^2}{(1-2e^t)^2}.
\end{align*}
Since $\psi(0)=1$, \eqref{eq:deriv} implies
\[ -\ln \psi(t)\geq \int_0^t \lambda z(k,u)\; \mathrm{d}u. \] 
Let $b\geq 0$. Since for $t\geq 0$ we have 
\[e^{-tb\lambda}\Pr[S\leq b\lambda]\leq E[e^{-tS}],\]
it follows that
\[ \ln \Pr[S\leq b\lambda]\leq -\lambda\int_0^t{z(k,u)}\mathrm{d}u +tb\lambda. \] 
Recall that in our setting we consider the value $\ln \Pr[S\leq pn]$ where $p$ is the constraint and $n$ is the length of the sequence. 
Therefore, since $\lambda=\frac{n}{2^{k+1}}$, we set $p=\frac{b}{2^{k+1}}$.
For any $k$, the following upper bound on the capacity holds for any $t\geq 0$, 
\begin{align*}
C_{k,p} &= 1+\limsup_{n\to\infty}\frac{1}{n}\log_2 \Pr[S\leq pn]\\
&\leq 1-\frac{\log_2 e}{2^{k+1}}\int_0^t{z(k,u)}\mathrm{d}u +tp\log_2 e,
\end{align*}
where we note that the change of logarithm base introduces a factor
of $\log_2 e$. Thus,
\begin{equation}
\label{eq:lb1}
1-C_{k,p}\geq \frac{\log_2 e}{2^{k+1}}\int_0^t{z(k,u)}\mathrm{d}u -tp\log_2 e.
\end{equation}

Note that for any $t\geq 0$, by Lebesgue's dominated convergence we
obtain that
\begin{align*}
\limup{k} \int_0^t z(k,u)\mathrm{d}u &= \int_0^t \limup{k} z(k,u)\mathrm{d}u\\
&=\int_0^t{\frac{e^u}{(1-2e^u)^2}}\mathrm{d}u.
\end{align*}
It follows that for any fixed $b > 0$, multiplying the right-hand side of
\eqref{eq:lb1} by $2^{k+2}/b$ gives,
\begin{align*} 
& \limup{k} \frac{2^{k+2}\log_2 e}{b}\parenv{\frac{1}{2^{k+1}}\int_0^t{z(k,u)}\mathrm{d}u -tp} \\ 
&\qquad = \frac{2\log_2 e}{b}\parenv{\int_0^t{\frac{e^u}{(1-2e^u)^2}}\mathrm{d}u -tc} \\
&\qquad =\frac{\log_2 e}{b}\parenv{1+\frac{1}{1-2e^t} -2tc}.
\end{align*}
Thus, to get a bound of the claimed form, this expression must equal $1$,
i.e.,
\[b=\parenv{1+\frac{1}{1-2e^t} -2tc}\log_2 e.\]
Since we are lower-bounding $1-C_{k,p}$, we would like to maximize
this expression by choosing an appropriate value of $t$. When $c>0$ the
maximum is attained by
\[ t= \ln \parenv{\frac{1+4c+\sqrt{1+8c}}{8c}}, \]
and then we get $b=\blo$, i.e.,
\[ b= \frac{3-\sqrt{1+8c}}{2}\log_2 e
-2c\log_2\parenv{\frac{1+4c+\sqrt{1+8c}}{8c}}.\] When $c=0$ we take
the limit as $t\to \infty$ to obtain $b=\blo=\log_2 e$, which completes the
proof.
\end{IEEEproof}

We note that taking $c=0$ in Theorem \ref{th:upasbnd} gives
$1-C_{k,0}\geq \frac{\log_2 e}{4\cdot 2^k}(1+o(1))$, which coincides with
the capacity's rate of convergence for the fully-constrained system
\cite{SchVar11}.

\subsection{A Lower Bound on the Capacity of $(0,k,p)$-RLL SCS}

We turn to consider a lower bound on the capacity of the $(0,k,p)$-RLL
SCS. We can extend the method of monotone families that was used in
\cite{SchVar11} to obtain such a bound. However, the result that we
describe next, which is based on the theory of large deviations,
outperforms the monotone-families approach.

As mentioned in \Tref{ldtheo}, the capacity of $(0,k,p)$-RLL SCS is
given by
\[ \ccap(\mathcal{F},p)=1-\inf_{\nu\in\Gamma} I(\nu)\] 
where $\Gamma=\mathset{\nu\in M_{\sigma}(\Sigma^k)~:~ \nu(1^{k+1})\leq p}$. 
The main idea of the bound is that by fixing some
$\nu\in\Gamma$ we find a lower bound on the capacity. We do, however,
have to keep in mind that the measure we choose must be
shift invariant.

\begin{theorem}
\label{th:lb}
For all $k\geq 1$ and $0\leq p\leq 2^{-(k+1)}$,
\begin{align}
C_{k,p} &\geq 1-\frac{1-p}{2^{k+1}-1}\log_2\parenv{\frac{2-2p}{1+2p(2^k-1)}}\nonumber \\ 
\label{eq:ub}
&\quad\  -p\log_2\parenv{\frac{2p(2^{k+1}-1)}{1+2p(2^k-1)}}.
\end{align}
\end{theorem}
\begin{IEEEproof}
Construct the following measure,
\[
\nu^*(i)=
\begin{cases}
p & i=1^{k+1}, \\
\frac{1-p}{2^{k+1} -1} & \text{otherwise.}
\end{cases}
\]
It is easy to verify that $\nu^*$ is indeed a shift-invariant measure.
Plugging $\nu^*$ into \Tref{ldtheo} gives,
\begin{align*}
C_{k,p} & = 1-\inf_{\nu\in\Gamma} I(\nu) \geq 1-I(\nu^*)\\
&= 1-\frac{1-p}{2^{k+1}-1}\log_2\parenv{\frac{2-2p}{1+2p(2^k-1)}} \\ 
&\quad\  -p\log_2\parenv{\frac{2p(2^{k+1}-1)}{1+2p(2^k-1)}},
\end{align*}
as claimed.
\end{IEEEproof}

The bound of Theorem \ref{th:lb} can now be used to prove an asymptotic
form when $k\to\infty$.

\begin{theorem}
  \label{th:lbasympt}
Let $c=\limup{k} \frac{p}{2^{-(k+1)}}$, where $c\in [0,1]$, and let
$\bup=(1+c)(1-H(\frac{1}{c+1}))$, where $H(\cdot)$ is the binary
entropy function. Then,
\[1-C_{k,p} \leq \frac{\bup}{2^{k+1}}(1+o(1)). \]
\end{theorem}
\begin{IEEEproof}
We take the limit of the right-hand side of \eqref{eq:ub} divided
by $\bup/2^{k+1}$. We obtain that 
\begin{align*} 
\limup{k}&\frac{2^{(k+1)}}{\bup}\left(\frac{(1-p)\log_2\parenv{\frac{2-2p}{1+2p(2^k-1)}} }{2^{k+1}-1}\right. \\ 
&\quad \left. +p\log_2\parenv{\frac{2p(2^{k+1}-1)}{1+2p(2^k-1)}}\right)\\
&= \frac{\log_2\parenv{\frac{2}{1+c}}}{\bup}+\frac{c}{\bup}\log_2\parenv{\frac{2c}{1+c}} \\
&= \frac{1}{\bup}\parenv{ (1+c)-(1+c)H\parenv{\frac{1}{1+c}} } \\
&=1,
\end{align*}
which proves the claim.
\end{IEEEproof}

In order to obtain a lower bound on the capacity of the
$D$-dimensional $(0,k,p)$-RLL SCS we use the method of monotone
families. The bound is recursive in the sense that it is given in
terms of the one-dimensional capacity. Thus, the expression may be
further simplified by plugging in lower bounds on the one-dimensional
capacity from \Tref{th:lb} or \Tref{th:lbasympt}. We follow the steps
presented in \cite{SchVar11}, and therefore, only sketch the
proof.

\begin{theorem}
The capacity of the $D$-dimensional $(0,k,p)$-RLL SCS is bounded by
the following,
\[C_{k,p}^{(D)} \geq 1+D\parenv{C_{k,p/D}^{(1)}-1}.\]
\end{theorem}

\begin{IEEEproof}[Sketch of proof]
Fix $j\in [D]$, and let $A_{j}$ denote the set of all
$\seq{\omega}\in\mathset{0,1}^{[n]^D}$ such that
$\seq{\omega}_{\seq{i},n,\seq{e}_j}$ are each one-dimensional
$(0,k,p/D)$-RLL semiconstrained strings. As in \cite{SchVar11}, we
note that the $D$-dimensional $(0,k,p)$-RLL SCS is a superset of the
intersection $\bigcap_{j\in [D]} A_j$. Additionally, each $A_j$ is a
monotone decreasing family in the sense that it is closed under the
operation of turning $1$'s into $0$'s. Thus, as in
\cite[Corollary 8]{SchVar11}, we obtain the desired result.
\end{IEEEproof}

%% Fig.~\ref{4plots} shows the upper bounds together with the lower
%% bounds for $(0,k,p)$-RLL when $k\in \mathset{1,2,3,4}$ as a function
%% of $p$.

%% \begin{figure*}[ht!]
%% \begin{center}
%% \begin{psfrags}
%%    \psfragscanon
%%    \psfrag{a}[][][0.9]{$(0,1,p)$-RLL, $0\leq p\leq 0.25$}
%%    \psfrag{b}[][][0.9]{$p$}
%%    \psfrag{c}[][][0.9]{$\ccap(p)$}
%%    \psfrag{d}[][][0.8]{$(0,2,p)$-RLL, $0\leq p\leq 0.125$}
%%    \psfrag{e}[][][0.9]{$p$}
%%    \psfrag{f}[][][0.9]{$(0,3,p)$-RLL, $0\leq p\leq 0.063$}
%%    \psfrag{g}[][][0.9]{$p$}
%%    \psfrag{h}[][][0.9]{$(0,4,p)$-RLL, $0\leq p\leq 0.0325$}
%%    \psfrag{j}[][][0.9]{$p$}

%% \includegraphics[width=15cm]{4plotsbounds.eps}
%% \caption{Upper and lower bounds as a function of $p$ for $(0,k,p)$-RLL where $k\in \mathset{1,2,3,4}$. }
%% \label{4plots}
%% \psfragscanoff
%% \end{psfrags}
%% \end{center}
%% \line(1,0){515}
%% \end{figure*}
%%%%%%%%%%%%%%%%%%%%%%%%%%%%%%%%%%%%%%%%%%%%%%%%%%%%%%%%%%%%%%%%%%%%%%%%

As a final comment we note that the ratio between the bounds of
\Tref{th:upasbnd} and \Tref{th:lbasympt} is at most $\approx 1.5$.

\section{Encoder and Decoder Construction for WSCS}
\label{sec:encdec}
%%%%%%%%%%%%%%%%%%%%%%%%%%%%%
%%%%%%%%%%%%%%%%%%%%%%%%%%%%%
%%%%%%%%%%%%%%%%%%%%%%%%%%%%%

In this section we describe an encoding and decoding scheme for
\emph{general} weak semiconstrained systems, that asymptotically
achieves capacity. The scheme relies on LD theory, and its
implementation is inspired by the coding scheme briefly sketched in
\cite{AviSieWol2005}.

We outline the strategy used to construct the encoder. Given a general
semiconstrained system, by LD theory we can solve an optimization
problem to find the empirical distribution of $k$-tuples that both
satisfies the semiconstraints, as well as maximizes the entropy. We
then use this empirical distribution to construct a Markov chain over
a De-Bruijn graph of order $k-1$, with a stationary distribution of
edges matching the empirical distribution given by LD theory.  We then
use this Markov chain to translate a stream of input symbols into
symbols that are sent over a channel. The decoder simply reverses the
process to obtain the input symbols.

The encoder we present is a block encoder which is also a constant bit
rate to constant bit rate encoder. We analyze it for input blocks that
contain i.i.d.~Bernoulli$(1/2)$ bits. In what follows we present some
notation, then describe the encoder and decoder, and finally, analyze
the scheme and show its rate is asymptotically optimal, and its
probability of failure tends to $0$.

\subsection{Preliminaries}

Several assumptions will be made in this section, all of them solely
for the purpose of simplicity of presentation. We will make these
assumptions clear. We further note that the results easily apply to
the general case as well.

Let $(\cF,\seq{P})$ be a WSCS. The first assumption
we make is that the system is over the binary alphabet
$\Sigma=\mathset{0,1}$. Another assumption we make is that
$\cF\subseteq \Sigma^k$, i.e., every word $\seq{\phi}\in\cF$ is of
the same length $k$ (see Theorem \ref{caplim2}).

Solving the appropriate LD problem (see Theorem \ref{ldtheo}) yields
the capacity of the system, which is denoted by
$C=\tcap(\cF,\seq{P})$, together with an optimal probability vector,
$\seq{p}$, of length $2^k$.  Each entry of the vector $\seq{p}$
corresponds to a $k$-tuple and contains the probability that a
$k$-tuple should appear in order to achieve the capacity of the
system, as well as satisfy the constraints.  We denote the entries
$\seq{p}=(p_0,p_1,\dots,p_{2^k-1})$.

Let $G$ be the binary De-Bruijn graph of order $k-1$, i.e., the vertices
are all the binary $(k-1)$-tuples, and the directed labeled edges are
\begin{equation}
  \label{eq:edge}
  \seq{u}=(u_1,u_2,\dots,u_{k-1}) \walk{u_k} (u_2,u_3,\dots,u_k)=\seq{u'},
\end{equation}
where $u_i\in\Sigma$. Thus, each vertex has $2$ outgoing edges labeled
$0$ and $1$. Additionally, each edge corresponds to a binary
$k$-tuple. For example, the edge from \eqref{eq:edge} corresponds to
$\seq{u}u_k = u_1 \seq{u'}$.

For convenience, we define an operator $\lchop:\Sigma^+\to\Sigma^*$,
(where $\Sigma^+$ denotes the set of positive-length finite strings
over $\Sigma$) which removes the first bit of a sequence. Namely, for
a sequence $\seq{u}=(u_1,u_2,\dots,u_n)\in\Sigma^n$, we define
$\lchop(\seq{u})=(u_2,u_3,\dots,u_n)\in\Sigma^{n-1}$. Thus, the edges
of the De-Bruijn graph are of the form $\seq{u}\to \lchop(\seq{u}a)$,
for all $\seq{u}\in\Sigma^{k-1}$ and $a\in\Sigma$. Another operator we
require is $\first:\Sigma^+\to\Sigma$, which maps to the first bit of
the sequence.  That is, $\first(\seq{u})=u_1$.

We can construct a Markov chain over $G$, whose transition matrix,
$A$, is a $2^{k-1}\times 2^{k-1}$ matrix whose $i,j$ entry, $A_{ij}$,
is the probability of choosing the edge going from vertex
$\seq{u}_i\in\Sigma^{k-1}$ to vertex $\seq{u}_j\in\Sigma^{k-1}$ given
that we are in state $\seq{u}_i$. At this point, for simplicity of
presentation, we assume that from each vertex emanate exactly two
outgoing edges with positive probability.

Denote by $\seq{v}=(v_0,v_1,\dots,v_{2^{k-1}-1})$ the stationary
distribution of the vertices of the Markov chain, i.e., $\seq{v}$ is
the unique left eigenvector of $A$ associated with the eigenvalue $1$,
whose entry sum is also $1$. We would like to find a Markov chain
on $G$ whose stationary distribution of the \emph{edges} matches
the vector $\seq{p}$. More precisely, the variables appear in the
non-zero entries of $A$ (we have $2^{k-1}$ variables), and we would
like to find a vector $\seq{v}$ as above (another set of $2^{k-1}$ variables)
satisfying
\begin{equation}
\label{eq:stat}
\seq{v}A=\seq{v},
\end{equation}
as well as, for each edge $\seq{u}_i\walk{a} \seq{u}_j$, $a\in\Sigma$,
\[v_i A_{i,j} = p_{\seq{u}_i a},\]
where $p_{\seq{u}_i a}$ is the entry in $\seq{p}$ that corresponds to
the $k$-tuple $\seq{u}_i a$.  We note that since the vector $\seq{p}$
is shift invariant, the set of equations has a solution (see
\cite{Hoc2012}).

\subsection{Encoder}

Assume $\seq{\omega}\in\Sigma^n$ is a sequence of $n$ input bits at
the encoder, which are i.i.d.~Bernoulli$(1/2)$. The encoding process
is comprised of three steps: partitioning, biasing, and graph walking.

\textbf{Partitioning:} The first step in the encoding process is
partitioning the sequence $\seq{\omega}$ of $n$ input bits into
$2^{k-1}$ subsequences of, perhaps, varying lengths, denoted $n_i$,
$0\leq i\leq 2^{k-1}-1$. Obviously, $n_i\geq 0$ for all $i$, as well
as $\sum_{i=0}^{2^{k-1}-1}n_i = n$. Each subsequence is to be
associated with a vertex of the Markov chain, or equivalently, with a
$(k-1)$-tuple. The first $n_0$ bits of the input are associated with
state $\seq{u}_0$, the following $n_1$ bits are associated with state
$\seq{u}_1$, and so on.

For every vertex $\seq{u}_i$, let $\seq{u}_j$ be the vertex for which
$\seq{u}_i\walk{0} \seq{u}_j$, i.e., $\seq{u}_j=\lchop(\seq{u}_i 0)$, and denote by $q_i$ the entry $A_{ij}$.
For every $k$-tuple $i$, let
\[\tilde{n_i}=H(q_i)v_i \cdot \frac{n}{C}.\]
For all $0\leq i\leq 2^{k-1}-1$ take $n_i=\fcenv{\tilde{n_i}}$, where
$\fcenv{\cdot}$ denotes either a rounding down or a rounding up.
The rounding is done in such a manner as to preserve the sum,
\[\sum_{i=0}^{2^{k-1}-1}\tilde{n_i}=\sum_{i=0}^{2^{k-1}-1} n_i.\]
This is always possible, for example, by taking $2^{k-1}$ steps, where
at the $i$th step, $\tilde{n_i}$ is rounded in a direction that keeps
the intermediate sum strictly less than $1$ away from the original
sum. We additionally note that indeed
\begin{align*}
\sum_{i=0}^{2^{k-1}-1} n_i &= \sum_{i=0}^{2^{k-1}-1}\tilde{n_i} \\
& \stackrel{(a)}{=} \frac{n}{C} \sum_{i\in\Sigma^{k-1}}H(q_i)v_i \\ 
&= \frac{n}{C} \sum_{i\in\Sigma^{k-1}}\left(-q_iv_i\log_2 q_i-(1-q_i)v_i\log_2(1-q_i)\right) \\ 
&= \frac{n}{C} \sum_{i\in\Sigma^{k-1}}\parenv{-p_{i0}\log_2 q_i -p_{i1}\log_2(1-q_i)} \\ 
&= \frac{n}{C} \sum_{i\in\Sigma^{k-1}} \left(-p_{i0}\log_2 \frac{p_{i0}}{p_{i0}+p_{i1}}\right. \\
&\quad \left. -p_{i1}\log_2 \frac{p_{i0}}{p_{i0}+p_{i1}}\right) \\ 
& \stackrel{(b)}{=} \frac{n}{C}\cdot C = n,
\end{align*}
where $(a)$ follows from the one-to-one correspondence between states
and $(k-1)$-tuples, and $(b)$ follows from Theorem \ref{ldtheo}.

\textbf{Biasing:} After obtaining $2^{k-1}$ subsequences, we take each
subsequence and bias it to create subsequences that are typical for a
Bernoulli$(q)$ source, for some $q$. To that end, we use an arithmetic
decoding process on each subsequence.

Let $\seq{\eta}_i$ be the subsequence that corresponds to vertex
$\seq{u}_i$, namely,
\[\seq{\eta}_i=\seq{\omega}_{\sum_{j=0}^{i-1}n_{j},n_i}.\]
For every $i$, we decode $\seq{\eta}_i$ using an arithmetic decoder with
probability $q_i$ to obtain a new sequence $\seq{\hat{\eta}}_i$ distributed
Bernoulli$(q_i)$.  Since the decoding process can continue
indefinitely, we stop the process when the obtained sequence
$\seq{\hat{\eta}}_i$ is of length
$\ceilenv{n_i/H(q_i)+n^{\frac{1}{2}+\epsilon}}$ bits for some known arbitrarily small $\epsilon\in (0,\frac{1}{4})$.  For every state
$\seq{u}_i$, we call the obtained sequence ``the information bits of state
$\seq{u}_i$.''

The resulting arithmetically-decoded sequence, $\seq{\hat{\eta}}_i$,
corresponds to a closed segment in $[0,1]$.  If there exists a state
$\seq{u}_i$ for which $\seq{\hat{\eta}}_i$ corresponds to a segment of
length greater than $2^{-n_i}$, an error is declared. For a detailed
description of arithmetic coding see \cite{Sai2004}.

\textbf{Graph walking:} The encoder now has the sequences
$\seq{\hat{\eta}}_i$, which are of various lengths. The encoder
appends to each sequence $\seq{\hat{\eta}}_i$ an extra
$\ceilenv{n^{\frac{1}{2}+2\epsilon}}$ bits distributed Bernoulli$(q_i)$. 
These extra bits carry no information and are used for padding only.

Then, the encoder starts the transmission in the following manner:
\begin{algorithm}[ht]
\caption{Encoding -- The Graph-Walking Stage}
\label{alg:enc}
\begin{algorithmic}
\State \textbf{Input:} The sequences $\seq{\hat{\eta}}_i$
\State \textbf{Output:} Transmitted bits
\State $\seq{u} \leftarrow 0^{k-1}$ \Comment{\textit{Set initial state}}
\Repeat
\If{$\seq{\hat{\eta}}_{\seq{u}}$ is an empty sequence}
\State Declare error and stop
\EndIf
\State $a \leftarrow \first(\seq{\hat{\eta}}_{\seq{u}})$ \Comment{\textit{Read first bit in queue}}
\State Transmit $a$
\State $\seq{\hat{\eta}}_{\seq{u}} \leftarrow \lchop(\seq{\hat{\eta}}_{\seq{u}})$ \Comment{\textit{Remove first bit from queue}}
\State $\seq{u} \leftarrow \lchop(\seq{u}a)$ \Comment{\textit{Proceed to the next state}}
\Until{$\ceilenv{\frac{n}{C}+n^{\frac{1}{2}+2\epsilon}}$ bits are transmitted}
\If{$\exists \seq{u}\in\Sigma^{k-1}$ s.t.~$\abs{\seq{\hat{\eta}}_{\seq{u}}} > \ceilenv{n^{\frac{1}{2}+2\epsilon}}$}
\State Declare error and stop
\EndIf
\end{algorithmic}
\end{algorithm}

Intuitively, when arriving at a state, the encoder takes a bit from
the sequence associated with the state. This bit is transmitted,
removed from the sequence, and determines the next state. The encoder
fails if a bit is required and the sequence associated with the state
is already empty, or if at the end of the main loop, not all
information bits have been transmitted.

\subsection{Decoder}
%%%%%%%%%%%%%%%%%%%%%
%%%%%%%%%%%%%%%%%%%%%%
%%%%%%%%%%%%%%%%%%%%%%
The decoding process mirrors the encoding. A simple graph walking
is the first stage of the decoding:
\begin{algorithm}[ht]
\caption{Decoding -- The Graph-Walking Stage}
\label{alg:dec}
\begin{algorithmic}
\State \textbf{Input:} Received bits
\State \textbf{Output:} The sequences $\seq{\hat{\eta}}_i$
\State $\seq{u} \leftarrow 0^{k-1}$ \Comment{\textit{Set initial state}}
\State Set $\seq{\hat{\eta}}_i$ to be empty sequences, for all $i$
\Repeat
\State Receive a bit $a$
\State $\seq{\hat{\eta}}_{\seq{u}} \leftarrow \seq{\hat{\eta}}_{\seq{u}}a$ \Comment{Append received bit to queue}
\State $\seq{u} \leftarrow \lchop(\seq{u}a)$ \Comment{\textit{Proceed to the next state}}
\Until{$\ceilenv{\frac{n}{C}+ n^{\frac{1}{2}+2\epsilon}}$ bits are received}
\end{algorithmic}
\end{algorithm}

After receiving the transmission, the decoder takes from each received
subsequence $\seq{\tilde{\eta}}_i$ only the first $\ceilenv{n_i
  /H(q_i)+n^{\frac{1}{2}+\epsilon}}$ bits and passes them through an
arithmetic encoder, thus reversing the second stage of the
encoder. The resulting sequences are now $\seq{\eta}_i$ of length
$n_i$. Finally, the decoder takes each $\seq{\eta}_i$ and concatenates
them in order to obtain the desired input sequence
$\seq{\omega}=\seq{\eta}_{0}\dots\seq{\eta}_{2^{k-1}-1}$.

\subsection{Analysis}
%%%%%%%%%%%%%%%%%%%%%%%%%%%%%%%%%%%%%%

We first show that the transmitted sequence indeed admits the
constraints given by $\seq{P}$.  Let $G$ be the De-Bruijn graph and
$A$ be the associated transition matrix with the stationary
distribution vector $\seq{v}$. It is easy to see that $G$ is
irreducible and aperiodic. It is well known that for such graphs,
starting with any vertex-probability vector $\seq{u}$, $\limup{n}
\seq{u}A^n=\seq{v}$.  For $\epsilon>0$, a divergence of $\epsilon$ in
some coordinate of $\seq{v}$ induces a divergence of $\epsilon$ in
some coordinate in $\seq{p}$.  Although the WSCS allows some
tolerance, we need to make sure that the tolerance is indeed $o(1)$.
To show that for large enough $n$ the transmitted words satisfy the
semiconstraints we need the following theorem.

\begin{theorem}[\cite[Ch. 4]{LevPerWil2006}]
Suppose $A$ is the transition matrix of an irreducible and
aperiodic Markov chain, with stationary distribution $\seq{v}$.  Then
there exist constants $\alpha\in (0,1)$ and $c>0$ such that
\[ \max_{i} \|(A^n)_{i,\cdot}-\seq{v}\|_{TV}\leq c\alpha^n, \]
where $(A^n)_{i,\cdot}$ denotes the $i$th row of $A^n$, and
$\|\cdot\|_{TV}$ denotes the total variation norm.
\end{theorem}

This implies that the rate of convergence to the stationary
distribution is exponential and as such, the divergence from the
semiconstraints decays as $o(1)$.

We now examine the rate of the presented coding scheme. The encoder
takes $n$ input bits and transmits
$\ceilenv{\frac{n}{C}+n^{\frac{1}{2}+2\epsilon}}$ bits over the
channel.  Since $\epsilon\in (0,\frac{1}{4})$, the asymptotic rate of
the scheme is
\[ \lim_{n\to\infty} \frac{n}{\ceilenv{\frac{n}{C}+n^{\frac{1}{2}+2\epsilon}}} = C,\]
and the coding scheme is asymptotically capacity achieving.

We now show that the error probability vanishes as $n$ grows. We
define the following events:
\begin{enumerate}
\item $E_1$: There exists a state $\seq{u}_i$ for which the arithmetic-decoded
  word $\seq{\hat{\eta}}_i$ corresponds to a segment of length greater
  than $2^{-n_i}$.
\item $E_2$: Some of the information bits have not been transmitted,
  i.e., there exists a state $j$ which, during the graph-walking
  stage, is visited strictly less than
  $\ceilenv{n_j/H(q_j)+n^{\frac{1}{2}+\epsilon}}$ times.
\item $E_3$: There exists a state $j$ which, during the graph-walking
  stage, is visited strictly more than
  $\ceilenv{n_j/H(q_j)+n^{\frac{1}{2}+\epsilon}+n^{\frac{1}{2}+2\epsilon}}$ times.
\end{enumerate}
Thus, the total error probability is 
\[ \perr=\Pr\sparenv{E_1\cup E_2\cup E_3}\leq \Pr\sparenv{E_1}+\Pr[E_2\cup E_3].\]
We bound the two probabilities appearing on the right-hand side
separately, showing each of them vanishes. 

We start by considering $\Pr[E_1]$. The arithmetic-coding scheme used
here receives a sequence of $n_i$ bits distributed Bernoulli$(1/2)$,
employs the \emph{decoding} process first, and then uses the encoding
process. The main obstacle in arithmetic decoding is that the
arithmetic decoder does not know when to stop the decoding process.
In our construction we stop the arithmetic decoder after
$\ceilenv{n_i/H(q_i)+n^{\frac{1}{2}+\epsilon}}$ bits are obtained. It is
well-known (for example, see \cite{Sai2004}) that the error probability in
the arithmetic-coding scheme vanishes as the block length grows, and
therefore, using a simple union bound $\Pr[E_1]$ tends to $0$.

We continue to the case of bounding $\Pr[E_2 \cup E_3]$. The encoder
transmits exactly $\ceilenv{n/C+n^{\frac{1}{2}+2\epsilon}}$ bits. Let $V$ be the
$2^{k-1}\times 2^{k-1}$ matrix all of whose rows are the stationary
vector $\seq{v}$ from \eqref{eq:stat}. We denote by $Z$ the matrix
\[Z=\parenv{I-A+V}^{-1},\]
where $A$ is from \eqref{eq:stat} and $I$ is the identity matrix. The
matrix $A$ is invertible by \cite[Chapter 11]{GriSne2006}. We also
define, for each $i$,
\[\sigma_i^2=2v_iZ_{ii}-v_i-v_i^2.\]
Let $S_i^{(n)}$ denote the number of times a walk of length $n$ on $G$
visits the vertex $i$.  Let us denote $f(n)=n^{\frac{1}{2}+\epsilon}$ and
$g(n)=n^{\frac{1}{2}+2\epsilon}$.  For any state $\seq{u}_i$,
it is easy to verify that since $v_i\neq 0,1$ for every $i$, 
\begin{align*} 
\limup{n} \frac{f(n)(1-v_i)-g(n)v_i}{\sqrt{\sigma_i^2\parenv{\frac{n}{C}+f(n)+g(n)}}}&\approx \limup{n} n^{\epsilon}(1-v_i(1+n^{\epsilon}))\\ 
&=-\infty
\end{align*} 
and that
\[ \limup{n} \frac{\parenv{f(n)+g(n)}(1-v_i)}{\sqrt{\sigma_i^2\parenv{\frac{n}{C}+f(n)+g(n)}}}\approx \limup{n} n^{2\epsilon}=\infty.\]

Using the central limit theorem (CLT) for Markov chains \cite[Chapter
  11]{GriSne2006} we bound $1-\Pr[E_2\cup E_3]$.  For any starting
vertex and for every state $i$, the probability that a walk of length
$\ceilenv{n/C+n^{\frac{1}{2}+\epsilon}+n^{\frac{1}{2}+2\epsilon}}$ on $G$ visits state $\seq{u}_i$ at
least $n_i/H(q_i)+n^{\frac{1}{2}+\epsilon}$ times but no more than
$\ceilenv{n_i/H(q_i)+n^{\frac{1}{2}+\epsilon}+n^{\frac{1}{2}+2\epsilon}}$ times is
given in (\ref{error}). Thus, as $n$ increases, the probability $\Pr[E_2\cup E_3]$ tends to $0$.

\begin{figure*}[t!]
\hrulefill
\begin{align}
\label{error} 
&\Pr\sparenv{\frac{n_i}{H(q_i)}+n^{\frac{1}{2}+\epsilon}<S_i^{(\frac{n}{C}+n^{\frac{1}{2}+\epsilon}+n^{\frac{1}{2}+2\epsilon})}<\frac{n_i}{H(q_i)}+n^{\frac{1}{2}+\epsilon}+n^{\frac{1}{2}+2\epsilon} } \nonumber \\
&\quad = \Pr\sparenv{\frac{nv_i}{C}+f(n)<S_i^{(\frac{n}{C}+f(n)+g(n))}< \frac{nv_i}{C}+f(n)+g(n)} \nonumber \\
&\quad = \Pr\left[\frac{\frac{nv_i}{C}+f(n)-\parenv{\frac{n}{C}+f(n)+g(n)}v_i}{\sqrt{\parenv{\frac{n}{C}+f(n)+g(n)}\sigma_i^2}}<\frac{S_i^{(\frac{n}{C}+f(n)+g(n))}-\parenv{\frac{n}{C}+f(n)+g(n)}v_i}{\sqrt{\parenv{\frac{n}{C}+f(n)+g(n)}\sigma_i^2}} \right. \nonumber \\
&\qquad \qquad \qquad \qquad \qquad \qquad \qquad \qquad \qquad \qquad \qquad \qquad \qquad 
< \left. \frac{\frac{nv_i}{C}+f(n)+g(n)-\parenv{\frac{n}{C}+f(n)+g(n)}v_i}{\sqrt{\parenv{\frac{n}{C}+f(n)+g(n)}\sigma_i^2}}\right] \nonumber \\ 
&\quad = \Pr \left[\frac{f(n)(1-v_i)-g(n)v_i}{\sqrt{\parenv{\frac{n}{C}+f(n)+g(n)}\sigma_i^2}}< 
\frac{S_i^{(\frac{n}{C}+f(n)+g(n))}-\parenv{\frac{n}{C}+f(n)+g(n)}v_i}{\sqrt{\parenv{\frac{n}{C}+f(n)+g(n)}\sigma_i^2}} < 
\frac{\parenv{f(n)+g(n)}(1-v_i)}{\sqrt{\parenv{\frac{n}{C}+f(n)+g(n)}\sigma_i^2}} \right] \nonumber \\
&\quad \xrightarrow[n\to\infty]{} \frac{1}{2\pi}\int_{-\infty}^{\infty} e^{-\frac{x^2}{2}}\mathrm{d}x =1
\end{align} 
\hrulefill
\end{figure*}

\section{Conclusion}
\label{sec:conc}
%%%%%%%%%%%%%%%%%%%%%%%%%%%%%%%%
%%%%%%%%%%%%%%%%%%%%%%%%%%%%%%%%
%%%%%%%%%%%%%%%%%%%%%%%%%%%%%%%%

In this paper we studied semiconstrained systems, as well as a
relaxation in the form of weak semiconstrained systems. We used tools
from probability theory, and in particular, large deviations theory,
to formulate closed-form bounds on the capacity of the $(0,k,p)$-RLL
SCS. These enabled us to bound the capacity's rate of convergence as
$k$ grows. We also examined WSCS and showed the limit in the
definition of the capacity for these systems does exist, unlike
SCS. We also showed the capacity is continuous, again, unlike the case
of SCS. Finally, we devised encoding and decoding schemes for WSCS
with rate that asymptotically achieves capacity, and with a vanishing
failure probability.

%% The
%% theory of large deviations was used to find the exact capacity as a
%% solution to a minimization problem. Other closed-form bounds on the
%% capacity were given as well. As a final somewhat simplistic motivating
%% example, assume a channel in which each transmitted $11$ may cause a
%% single bit error with probability $0.1$. Removing all $11$'s, i.e.,
%% using the inverted $(0,1)$-RLL, results in capacity $\approx
%% 0.6942$. Another option is using an error-correcting code, reaching a
%% capacity of $\approx 0.8313$.  Finally, we consider a combined scheme:
%% we reduce the frequency $11$ shows up in the channel to $p=0.141$ (but
%% not eliminate it entirely), by using $(0,1,p)$-RLL SCS, which has
%% capacity $\approx 0.9674$. We are then left with a BSC channel with
%% bit-error probability $0.1\cdot p=0.0141$, which has capacity
%% $1-H(0.0141)\approx 0.8931$. The total capacity of this combined
%% scheme is therefore $\approx 0.8640$. This comparison will be further
%% explored in the full version of the paper.

%% Other results, to be described in a fuller version, include a more
%% general treatment that is not limited to offending substrings of same
%% length, encoding schemes, and connections with the theory of formal
%% languages.

\section{Appendix}
%%%%%%%%%%%%%%%%%%%
%%%%%%%%%%%%%%%%%%%
We provide a proof that for every $n$, the set $\Gamma_n$ contains a probability measure which belongs to the support 
of $\mu_n$. 
\begin{definition}
Let $G=(V,E)$ be a directed graph. An \emph{$n$-circulation} is an
assignment of weights $w(\cdot)$ to the edges such that:
\begin{enumerate}
\item
  $w(e)\geq 0$ for all $e\in E$.
\item
  $\sum_{e\in \inc(v)}w(e) = \sum_{e\in \out(v)}w(e)$ for all $v\in V$.
\item
  $\sum_{e\in E} w(e) = n$.
\end{enumerate}
An \emph{integer $n$-circulation} is an $n$-circulation for which
$w(e)\in\Z$ for all $e\in E$.
\end{definition}

We assume throughout that a directed graph has no parallel edges.

\begin{definition}
Let $G=(V,E)$ be a directed graph. A cycle is a sequence
$v_0,v_1,\dots,v_{k-1}$, such that $v_i\in V$, and $(v_i,v_{i+1})\in
E$ for all $i$, where the indices are taken modulo $k$. The cycle is
\emph{vertex simple} if the vertices are all distinct. It is
\emph{edge simple} if the edges are all distinct.
\end{definition}

We note that using this notation, a cycle of length $1$ is described
by a sequence with one vertex only. We say two cycles are distinct if
they do not contain the exact same set of edges.

The underlying graph of a direct graph, is the undirected graph
obtained by removing the orientation of the edges. An underlying graph
may contain parallel edges.

\begin{definition}
Let $G=(V,E)$ be a directed graph. Let $C=v_0,\dots,v_{k-1}$ be a cycle
in the underlying graph. For all $i$, we say $(v_i,v_{i+1})$ is
cooriented if $(v_i,v_{i+1})\in E$, and disoriented if
$(v_{i+1},v_i)\in E$. The set of cooriented and disoriented edges
are defined as:
\begin{align*}
  \coo(C) & = \mathset{(v_i,v_{i+1}) : (v_i,v_{i+1})\in E},\\
  \doo(C) & = \mathset{(v_i,v_{i+1}) : (v_i,v_{i+1})\not\in E}.
\end{align*}
We say the \emph{effective length} of the cycle is $\coo(C)-\doo(C)$.
\end{definition}

A cycle with effective length of $0$ is called \emph{balanced}.

\begin{definition}
Let $G=(V,E)$ be a directed graph, $w$ a weight assignment to
the edges, and $C=v_0,\dots,v_{k-1}$ an edge-simple cycle in the
underlying graph. Let $\epsilon\in\R$. An $\epsilon$-adjustment of the
cycle $C$ is a weight assignment $w'$ such that,
\[ w'(e) = \begin{cases}
w(e)+\epsilon & e\in\coo(C),\\
w(e)-\epsilon & e\in\doo(C),\\
w(e) & \text{otherwise.}
\end{cases}\]
\end{definition}

\begin{lemma}
\label{lem:adj1}
Let $G=(V,E)$ be a directed graph, and let $C$ be a balanced
edge-simple cycle in the underlying graph. Assume $w$ is an
$n$-circulation, and $\epsilon\in\R$ is some real number. Denote by
$w'$ the edge-weighing function obtained from $w$ by
$\epsilon$-adjusting $C$. If $w'(e)\geq 0$ for all $e\in E$, then $w'$
is also an $n$-circulation.
\end{lemma}
\begin{IEEEproof}
Property 1 is satisfied by requirement. It is easily verifiable that
an adjustment preserves property 2. Finally, the overall weight of the
edges is not changed.
\end{IEEEproof}

\begin{lemma}
\label{lem:adj2}
Let $G=(V,E)$ be a directed graph, and let $C_1$ and $C_2$ be two
distinct edge-simple cycles in the underlying graph of effective
lengths $k_1$ and $k_2$ respectively. Assume $w$ is an
$n$-circulation, and $\epsilon\in\R$ is some real number. Denote by
$w'$ the edge-weighing function obtained from $w$ by
$k_2\epsilon$-adjusting $C_1$, and then $-k_1\epsilon$-adjusting
$C_2$. If $w'(e)\geq 0$ for all $e\in E$, then $w'$ is also an
$n$-circulation.
\end{lemma}
\begin{IEEEproof}
Property 1 is satisfied by requirement. It is easily verifiable that
an adjustment preserves property 2. Finally, the overall weight of the
edges if increased by $k_1k_2\epsilon$ after the first adjustment, and
decreased by the same amount after the second adjustment.
\end{IEEEproof}

\begin{theorem}
  \label{th:maindb}
Let $G=(V,E)$ be the De-Bruijn graph of order $m$ over the finite
alphabet $\Sigma$. Assume $w$ is an $n$-circulation for some
$n\in\N$. Then there exists an integer $n$-circulation $w'$ such that
\[\floorenv{w(e)}\leq w'(e) \leq \ceilenv{w(e)}+1,\]
for all $e\in E$.
\end{theorem}
\begin{IEEEproof}
We first look at the underlying unoriented graph. This is a regular
graph of degree $2\abs{\Sigma}$. Because of property 2, there is no
vertex with exactly one incident edge of non-integer weight. It
follows, that every edge of non-integer weight is on an unoriented
cycle in the underlying graph, all of whose edges have non-integer
weights. We call such cycles, \emph{non-integer cycles}.

Assume there is a balanced non-integer cycle $C$. By Lemma
\ref{lem:adj1}, and since all of the weights on the cycle's edges are
non-integers, there exists a minimal $\epsilon>0$ such that
$\epsilon$-adjusting $C$ creates a new $n$-circulation with at least
one edge of the cycle having an integer weight. Furthermore, for this
edge $e$, since we took the minimal $\epsilon$ possible, the new
weight of the edge is either $\floorenv{w(e)}$ or $\ceilenv{w(e)}$.

We can repeat the process as long as we have balanced non-integer
cycles. If we do not, assume we have two distinct non-integer cycles,
$C_1$ and $C_2$. We can assume they are edge simple. Again, there
exists a minimal $\epsilon>0$ such that adjusting by Lemma
\ref{lem:adj2} turns at least one of the cycle-edge weights to an
integer weight. Like before, choosing the minimal such $\epsilon$
ensures the new weight is either a rounding down or a rounding up of
the original weight.

After this, we go back to looking for balanced non-integer cycles, an
continue this way. Repeating the above, we must end up with either an
integer $n$-circulation $w'$ as desired, or with a circulation all of
whose non-integer weights form a single vertex-simple non-balanced
non-integer cycle. Denote this cycle as $C$, and assume it has an
effective length of $k$. It is easy to verify that the fractional part
of the weight of all cooriented edges is equal to some constant
$0<\alpha<1$, whereas the fractional part of the disoriented edges is
$1-\alpha$. Since the sum of the edges of $C$ is an integer, we have
\[\abs{\coo(C)}\alpha + \abs{\doo(C)}(1-\alpha)=\abs{\doo(C)}+k\alpha,\]
is an integer. Thus, $k\alpha$ is an integer and $0< k\alpha < k$.

It is well known \cite{Lem71} that the De Bruijn graph of order $m$
has an edge-simple directed cycle for each length between $1$ and
$\abs{\Sigma}^m$. We find such a cycle of length $k\alpha$. We then
round down all the weights of the cycle $C$, and add $1$ to all the
edges of the $k\alpha$-cycle. We call the resulting $n$-circulation
$w'$. Since the weights of the edges of the $k\alpha$ cycle may have
already been increased in a previous rounding operation, we have
\[\floorenv{w(e)}\leq w'(e) \leq \ceilenv{w(e)}+1,\]
for all $e\in E$, as claimed.
\end{IEEEproof}

Finally, in the proof of \Tref{caplim} we are given $q_1$, a
shift-invariant distribution over $\Sigma^k$. By identifying the
elements of $\Sigma^k$ with the edges of the De Bruijn graph of order
$k-1$ over $\Sigma$, and assigning each edge $\seq{\phi}\in\Sigma^k$
the weight $n\cdot q_1(\seq{\phi})$, we obtain a circulation. Using
Theorem \ref{th:maindb}, we can obtain an integer circulation, which
we denote $w'$. If we define $\omega_1=w'/n$, then $\omega_1$
is a shift-invariant distribution in $L_{n,k}^{\iseq{S}}$ satisfying
\[\|q_1-\omega_1\|_{\infty}\leq \frac{2}{n},\]
as claimed.

%% \section{Going to Appendix}
%% %%%%%%%%%%%%%%%%%%%%%%%%%%%%%
%% %%%%%%%%%%%%%%%%%%%%%%%%%%%%%
%% %%%%%%%%%%%%%%%%%%%%%%%%%%%%%

%% This upper bound can be slightly improved. 
%% In order to do so, we discard the use of Jensen's inequality in the proof of Theorem \ref{janson}.
%% We calculate the expression $\sum_A p_A E[e^{-tS_A'}|I_A=1]$, which is equal to $\frac{n}{16}\parenv{e^{-t}+2e^{-2t}+e^{-3t}}$.
%% Obtaining $\log_2{\Pr(S\leq pn)}\leq -\frac{n}{16}\parenv{\frac{7}{3}-e^{-t}-e^{-2t}-\frac{1}{3}e^{-3t}}+tpn$.
%% The $t$ which minimize the right-hand side is calculated as follows.
%% Denoting $t=\log_2{(x^{-1})}$, reduce the problem to find the roots of the polynomial $x^3+2x^2+x-16p$. 
%% The root which minimizes the polynomial is given by
%% \begin{equation}
%% x=\frac{(-1+(1+216p+12\sqrt{3p(1+108p)})^{\frac{1}{3}})^2}{3(1+216p+12\sqrt{3p(1+108p)})^{\frac{1}{3}}}.
%% \end{equation} 
%% And we have that 
%% \begin{equation}
%% \log_2 \Pr(S\leq pn)\leq -\frac{n}{16}\parenv{\frac{7}{3}-x-x^2-\frac{1}{3}x^3}+pn\log_2{x^{-1}}.
%% \end{equation}
%% The improvement is mostly at the area close to $0$.

%%%%%%%%%%%%%%%%%%%%%%%%%%%%%%%%%%%%%%%%%%%%%%%%%%%%%%%%%%%%%%%%%%%%%%%%%%%%%%%%%%%%%%
\bibliographystyle{IEEEtranS}
\bibliography{allbib}

\end{document}